\documentclass[12pt]{article}
\usepackage{amssymb}
\begin{document}



\def\a{\alpha}
\def\b{\beta}
\def\d{\delta}
\def\e{\epsilon}
\def\g{\gamma}
\def\h{\mathfrak{h}}
\def\k{\kappa}
\def\l{\lambda}
\def\o{\omega}
\def\p{\wp}
\def\r{\rho}
\def\t{t}
\def\s{\sigma}
\def\z{\zeta}
\def\x{\xi}
 \def\A{{\cal{A}}}
 \def\B{{\cal{B}}}
 \def\C{{\cal{C}}}
 \def\D{{\cal{D}}}
\def\G{\Gamma}
\def\K{{\cal{K}}}
\def\O{\Omega}
\def\R{\bar{R}}
\def\T{{\cal{T}}}
\def\L{\Lambda}
\def\f{E_{\tau,\eta}(sl_2)}
\def\E{E_{\tau,\eta}(sl_n)}
\def\Zb{\mathbb{Z}}
\def\Cb{\mathbb{C}}

\def\R{\overline{R}}

\def\beq{\begin{equation}}
\def\eeq{\end{equation}}
\def\bea{\begin{eqnarray}}
\def\eea{\end{eqnarray}}
\def\ba{\begin{array}}
\def\ea{\end{array}}
\def\no{\nonumber}
\def\le{\langle}
\def\re{\rangle}
\def\lt{\left}
\def\rt{\right}

\newtheorem{Theorem}{Theorem}
\newtheorem{Definition}{Definition}
\newtheorem{Proposition}{Proposition}
\newtheorem{Lemma}{Lemma}
\newtheorem{Corollary}{Corollary}
\newcommand{\proof}[1]{{\bf Proof. }
        #1\begin{flushright}$\Box$\end{flushright}}

\baselineskip=20pt

\newfont{\elevenmib}{cmmib10 scaled\magstep1}
\newcommand{\preprint}{
   \begin{flushleft}
   \end{flushleft}\vspace{-1.3cm}
   \begin{flushright}\normalsize
   \end{flushright}}
\newcommand{\Title}[1]{{\baselineskip=26pt
   \begin{center} \Large \bf #1 \\ \ \\ \end{center}}}
\newcommand{\Author}{\begin{center}
   \large \bf
Wen-Li Yang${}^{a}$, ~Xi Chen${}^{a}$, ~Jun Feng${}^{a}$,~Kun Hao${}^{a}$,~Ke Wu ${}^b$,
~Zhan-Ying Yang${}^c$ ~and~Yao-Zhong Zhang ${}^d$
 \end{center}}
\newcommand{\Address}{\begin{center}

     ${}^a$ Institute of Modern Physics, Northwest University,
     Xian 710069, P.R. China\\
     ${}^b$ School of Mathematical Science, Capital Normal University,
     Beijing 100037, P.R. China \\
     ${}^c$ The Department of Physics, Northwest University,
     Xian 710069, P.R. China \\
     ${}^d$ The University of Queensland, School of Mathematics and Physics,  Brisbane, QLD 4072,
     Australia\\
   \end{center}}
\newcommand{\Accepted}[1]{\begin{center}
   {\large \sf #1}\\ \vspace{1mm}{\small \sf Accepted for Publication}
   \end{center}}

\preprint
\thispagestyle{empty}
\bigskip\bigskip\bigskip

\Title{Drinfeld twist and symmetric Bethe vectors of  the open XYZ chain with non-diagonal
boundary terms } \Author

\Address
\vspace{1cm}

\begin{abstract}
With the help of the Drinfeld twist or factorizing
F-matrix for the  eight-vertex solid-on-solid (SOS) model, we find that
in the F-basis provided by the twist the two sets of pseudo-particle creation
operators simultaneously take completely symmetric and
polarization free form. This allows us to obtain the explicit and completely symmetric
expressions of the two sets of Bethe states of the model.

\vspace{1truecm} \noindent {\it PACS:} 03.65.Fd; 04.20.Jb;
05.30.-d; 75.10.Jm

\end{abstract}
\newpage
\section{Introduction}
\label{intro} \setcounter{equation}{0}

The algebraic Bethe ansatz \cite{Kor93} has been proven to provide
a powerful tool of solving eigenvalue problems of quantum
integrable systems such as quantum spin chains. In this framework,
the pseudo-particle creation and annihilation operators are
constructed by off-diagonal matrix elements of the so-called
monodromy matrix. The Bethe states (eigenstates) of transfer
matrix are obtained by applying creation operators to the
reference state (or pseudo-vacuum state). However, the apparently
simple action of creation operators is plagued with non-local
effects arising from polarization clouds or compensating exchange
terms on the level of local operators. This make the explicit
construction of the Bethe states challenging.

Progress for obtaining  explicit expressions of the Bethe states has
been made for the XXX and XXZ spin chains with periodic  boundary conditions
(or the closed XXX and XXZ chains) \cite{Mai00}  by using the so-called F-basis
provided by the Drinfeld twist or  factorizing F-matrix \cite{Dri83}.
In the F-basis, the pseudo-particle creation and annihilation
operators  of the models  take completely symmetric forms and
contain no compensating exchange terms on the level of local
operators (i.e. polarization free). As a result, the Bethe states
of the models are simplified dramatically and can be written down
explicitly \cite{Kit99}. Similar results have been obtained for
other models with periodic boundary conditions
\cite{Alb00-1,Alb00,Alb01,Yan06-1,Zha06}.

It was  shown \cite{Wan02,Kit07} that the F-matrices of
the closed XXX and XXZ chains also  make the pseudo-particle
creation operators of the open XXX and XXZ chains with diagonal
boundary terms \cite{Skl88} polarization free. This is mainly due to the fact
that the closed chain and the corresponding open chain with
diagonal boundary terms share the same reference state. However, the story for the open XXZ chain with
non-diagonal boundary terms is quite different
\cite{Nep04,Cao03,Yan04,Yan04-1,Gie05,Gie05-1,Gal05,Yan05,Baj06,Doi06,Mur06,Yan06,Bas07,Gal08,Mur09,Ami10,Cra10}.
Firstly, the reference state (all spin up state) of the closed
chain is no longer a reference state of the open chain with
non-diagonal boundary terms \cite{Cao03,Yan04,Yan04-1}. Secondly,
at least two reference states (and thus two sets of Bethe states) are
needed \cite{Yan07} for the open XXZ chain with non-diagonal
boundary terms in order to obtain its complete spectrum
\cite{Nep03,Yan06}. As a consequence, the F-matrix found in
\cite{Mai00} is no longer the {\it desirable} F-matrix for the
open XXZ chain with non-diagonal boundary terms. Recently, we have succeeded
in obtaining the factorizing F-matrices for the open XXZ chain with a non-diagonal
boundary terms \cite{Yan10} and using the F-matrices to construct the determinant representations
of the DW partition function of the six-vertex model with a non-diagonal reflection end \cite{Fil10,Yan10-1} and
the scalar products of the Bethe states of the open XXZ chain \cite{Yan11-1}.

In this paper, we focus on the most general Heisenberg spin chain---the
open XYZ chain \cite{Bax71,Bax82} with a non-diagonal boundary terms whose trigonometric/rational limit gives the open
XXZ/XXX chain. With the help of the F-matrices of the eight-vertex SOS model \cite{Alb00}, we find that in the
F-basis the two sets of pseudo-particle creation operators (acting
on the two reference states) of the boundary model simultaneously
take completely symmetric and polarization free forms. These
enable us to derive the explicit and completely symmetric
expressions of the two sets of Bethe states of the model. Moreover, the coefficients in these  expressions
can be expressed in terms of a single determinant. Such a single determinant representation will be essential for the study
of the scalar products of the Bethe vectors of the open XYZ chain with non-diagonal boundary terms.

The paper is organized as follows.  In section 2, we briefly
describe the open XYZ chain with non-diagonal boundary terms and
introduce the pseudo-particle creation operators and the two sets
of Bethe states of the model. In section 3, we introduce the face picture
of the model and express the two sets of Bethe states in terms of their
face-picture versions. In section 4, with the help of the F-matrix of the
eight-vertex SOS model, we obtain   the completely
symmetric and polarization free representations of the
pseudo-particle creation operators. In section 5, we give the complete symmetric
expressions of the two sets of Bethe states in the F-basis, in which the coefficients
can be expressed in terms of a single determinant  respectively.
In section 6, we summarize our results and give some discussions.


\section{ The inhomogeneous spin-$\frac{1}{2}$ XYZ open chain}
\label{XYZ} \setcounter{equation}{0}

Let us fix $\tau$ such that ${\rm Im}(\tau)>0$ and a generic complex number $\eta$.
Introduce the following elliptic functions
\bea
\theta\lt[\begin{array}{c} a\\b
  \end{array}\rt](u,\tau)&=&\sum_{n=-\infty}^{\infty}
  \exp\lt\{i\pi\lt[(n+a)^2\tau+2(n+a)(u+b)\rt]\rt\},\label{Function-a-b}\\
\theta^{(j)}(u)&=&\theta\lt[\begin{array}{c}\frac{1}{2}-\frac{j}{2}\\
 [2pt]\frac{1}{2}
 \end{array}\rt](u,2\tau),\quad j=1,2;\qquad
 \s(u)=\theta\lt[\begin{array}{c}\frac{1}{2}\\[2pt]\frac{1}{2}
 \end{array}\rt](u,\tau).
 \label{Function-j}\eea The
$\s$-function\footnote{Our $\s$-function is the
$\vartheta$-function $\vartheta_1(u)$ \cite{Whi50}. It has the
following relation with the {\it Weierstrassian\/} $\s$-function
$\s_w(u)$: $\s_w(u)\propto e^{\eta_1u^2}\s(u)$ with
$\eta_1=\pi^2(\frac{1}{6}-4\sum_{n=1}^{\infty}\frac{nq^{2n}}{1-q^{2n}})
$ and $q=e^{i\tau}$.}
 satisfies the so-called Riemann
identity:\bea
&&\s(u+x)\s(u-x)\s(v+y)\s(v-y)-\s(u+y)\s(u-y)\s(v+x)\s(v-x)\no\\
&&~~~~~~=\s(u+v)\s(u-v)\s(x+y)\s(x-y),\label{identity}\eea which
will be useful in the following. Moreover, for any $\a=(\a_1,\a_2),\,\a_1,\a_2\in\Zb_2$, we
can introduce a function $\s_{\a}(u)$ as follow
\bea
  \s_{\a}(u)=\theta\lt[\begin{array}{c}\frac{1}{2}+\frac{\a_1}{2}\\[2pt]\frac{1}{2}+\frac{\a_2}{2}
 \end{array}\rt](u,\tau),\quad\quad \a_1,\a_2\in \Zb_2.\label{sigma-function}
\eea The above definition implies the identification
  $\s_{(0,0)}(u)=\s(u)$.

Let $V$ be a two-dimensional vector space $\Cb^2$ and
$\{\e_i|i=1,2\}$ be the orthonormal basis  of $V$ such that
$\langle \e_i,\e_j\rangle=\d_{ij}$. The well-known eight-vertex
model R-matrix $\R(u)\in {\rm End}(V\otimes V)$ is given by \bea
\R(u)=\lt(\begin{array}{llll}a(u)&&&d(u)\\&b(u)&c(u)&\\
&c(u)&b(u)&\\d(u)&&&a(u)\end{array}\rt). \label{r-matrix}\eea The
non-vanishing matrix elements  are \cite{Bax82}\bea
&&a(u)=\frac{\theta^{(1)}(u)\,\theta^{(0)}(u+\eta)\,\s(\eta)}
{\theta^{(1)}(0)\, \theta^{(0)}(\eta)\,\s(u+\eta)},\quad
b(u)=\frac{\theta^{(0)}(u)\,
 \theta^{(1)}(u+\eta)\,\s(\eta)}
{\theta^{(1)}(0)\,\theta^{(0)}(\eta)\,\s(u+\eta)},\no\\[6pt]
&&c(u)=\frac{\theta^{(1)}(u)\,
 \theta^{(1)}(u+\eta)\,\s(\eta)}
{\theta^{(1)}(0)\, \theta^{(1)}(\eta)\,\s(u+\eta)},\quad
d(u)=\frac{\theta^{(0)}(u)\,
 \theta^{(0)}(u+\eta)\,\s(\eta)}
{\theta^{(1)}(0)\theta^{(1)}(\eta)\,\s(u+\eta)}.\label{r-func}\eea Here
$u$ is the spectral parameter and $\eta$ is the so-called crossing
parameter. The R-matrix satisfies the quantum Yang-Baxter equation
(QYBE)
\bea \R_{1,2}(u_1-u_2)\R_{1,3}(u_1-u_3)\R_{2,3}(u_2-u_3)
=\R_{2,3}(u_2-u_3)\R_{1,3}(u_1-u_3)\R_{1,2}(u_1-u_2).\label{QYBE}\eea
Throughout we adopt the standard notation: for any
matrix $A\in {\rm End}(V)$, $A_j$ (or  $A^j$)is an embedding operator in the
tensor space $V\otimes V\otimes\cdots$, which acts as $A$ on the
$j$-th space and as identity on the other factor spaces;
$R_{i,j}(u)$ is an embedding operator of R-matrix in the tensor
space, which acts as identity on the factor spaces except for the
$i$-th and $j$-th ones.

One introduces the ``row-to-row"  (or one-row ) monodromy matrix
$T(u)$, which is an $2\times 2$ matrix with elements being
operators acting on $V^{\otimes N}$, where $N=2M$ ($M$ being a
positive integer),\bea
T_0(u)=\R_{0,N}(u-z_N)\R_{0,N-1}(u-z_{N-1})\cdots
\R_{0,1}(u-z_1).\label{Mon-V}\eea Here $\{z_j|j=1,\cdots,N\}$ are
arbitrary free complex parameters which are usually called
inhomogeneous parameters.

Integrable open chain can be constructed as follows \cite{Skl88}.
Let us introduce a pair of K-matrices $K^-(u)$ and $K^+(u)$. The
former satisfies the reflection equation (RE)
 \bea &&\R_{1,2}(u_1-u_2)K^-_1(u_1)\R_{2,1}(u_1+u_2)K^-_2(u_2)\no\\
 &&~~~~~~=
K^-_2(u_2)\R_{1,2}(u_1+u_2)K^-_1(u_1)\R_{2,1}(u_1-u_2),\label{RE-V}\eea
and the latter  satisfies the dual RE \bea
&&\R_{1,2}(u_2-u_1)K^+_1(u_1)\R_{2,1}(-u_1-u_2-2\eta)K^+_2(u_2)\no\\
&&~~~~~~=
K^+_2(u_2)\R_{1,2}(-u_1-u_2-2\eta)K^+_1(u_1)\R_{2,1}(u_2-u_1).
\label{DRE-V}\eea For open spin-chains, instead of the standard
``row-to-row" monodromy matrix $T(u)$ (\ref{Mon-V}), one needs to
consider  the
 ``double-row" monodromy matrix $\mathbb{T}(u)$
\bea
  \mathbb{T}(u)=T(u)K^-(u)\hat{T}(u),\quad \hat{T}(u)=T^{-1}(-u).
  \label{Mon-V-0}
\eea Then the double-row transfer matrix of the XYZ chain with
open boundary (or the open XYZ chain) is given by
\bea
  \t(u)=tr(K^+(u)\mathbb{T}(u)).\label{trans}
\eea The QYBE and
(dual) REs lead to that the transfer matrices with different
spectral parameters commute with each other \cite{Skl88}:
$[\t(u),\t(v)]=0$. This ensures the integrability of the open XYZ
chain.

In this paper, we consider the K-matrix $K^{-}(u)$ which is a
generic solution \cite{Ina94,Hou95} to the RE (\ref{RE-V}) associated with the  R-matrix
(\ref{r-matrix})
\bea K^-(u)=k^-_0(u)+k^-_x(u)\,\s^x+k^-_y\,\s^y+k^-_z(u)\,\s^z,\label{K-matrix}\eea
where $\s^x,\s^y,\s^z$ are the Pauli matrices and the coefficient functions are
\bea
&& k^-_0(u)=\frac{\s(2u)\,\s(\l_1+\l_2-\frac{1}{2})\,\s(\l_1+\xi)\,\s(\l_2+\xi)}
   {2\,\s(u)\,\s(-u+\l_1+\l_2-\frac{1}{2})\,\s(\l_1+\xi+u)\,\s(\l_2+\xi+u)},\no\\[6pt]
&& k^-_x(u)=\frac{\s(2u)\,\s_{(1,0)}(\l_1+\l_2-\frac{1}{2})\,\s_{(1,0)}(\l_1+\xi)\,\s_{(1,0)}(\l_2+\xi)}
   {2\,\s_{(1,0)}(u)\,\s(-u+\l_1+\l_2-\frac{1}{2})\,\s(\l_1+\xi+u)\,\s(\l_2+\xi+u)},\no\\[6pt]
&& k^-_y(u)=\frac{i\,\s(2u)\,\s_{(1,1)}(\l_1+\l_2-\frac{1}{2})\,\s_{(1,1)}(\l_1+\xi)\,\s_{(1,1)}(\l_2+\xi)}
   {2\,\s_{(1,1)}(u)\,\s(-u+\l_1+\l_2-\frac{1}{2})\,\s(\l_1+\xi+u)\,\s(\l_2+\xi+u)},\no\\[6pt]
&& k^-_z(u)=\frac{\s(2u)\,\s_{(0,1)}(\l_1+\l_2-\frac{1}{2})\,\s_{(0,1)}(\l_1+\xi)\,\s_{(0,1)}(\l_2+\xi)}
   {2\,\s_{(0,1)}(u)\,\s(-u+\l_1+\l_2-\frac{1}{2})\,\s(\l_1+\xi+u)\,\s(\l_2+\xi+u)}.\label{K-matrix-2-1} \eea
At the same time, we introduce  the corresponding {\it dual\/}
K-matrix $K^+(u)$ which is a generic solution to the dual
reflection equation (\ref{DRE-V}) with a particular choice of the
free boundary parameters according to those of $K^-(u)$ (\ref{K-matrix})-(\ref{K-matrix-2-1}):
\bea
 K^+(u)=k^+_0(u)+k^+_x(u)\,\s^x+k^+_y\,\s^y+k^+_z(u)\,\s^z,
 \label{DK-matrix}
\eea with the coefficient functions
\bea
&& k^+_0(u)=\frac{\s(\hspace{-0.06truecm}-\hspace{-0.06truecm}2u\hspace{-0.06truecm}-\hspace{-0.06truecm}2\eta)
   \s(\l_1\hspace{-0.06truecm}+\hspace{-0.06truecm}\l_2\hspace{-0.06truecm}+\hspace{-0.06truecm}\eta
   \hspace{-0.06truecm}-\hspace{-0.06truecm}\frac{1}{2})
   \s(\l_1\hspace{-0.06truecm}+\hspace{-0.06truecm}\bar{\xi})
   \s(\l_2\hspace{-0.06truecm}+\hspace{-0.06truecm}\bar{\xi})}
   {2\s(\hspace{-0.06truecm}-\hspace{-0.06truecm}u\hspace{-0.06truecm}-\hspace{-0.06truecm}\eta)
   \s(u\hspace{-0.06truecm}+\hspace{-0.06truecm}\eta\hspace{-0.06truecm}+\hspace{-0.06truecm}\l_1\hspace{-0.06truecm}+\hspace{-0.06truecm}\l_2
   \hspace{-0.06truecm}-\hspace{-0.06truecm}\frac{1}{2})\s(\l_1\hspace{-0.06truecm}+\hspace{-0.06truecm}\bar{\xi}
   \hspace{-0.06truecm}-\hspace{-0.06truecm}u\hspace{-0.06truecm}-\hspace{-0.06truecm}\eta)
   \s(\l_2\hspace{-0.06truecm}+\hspace{-0.06truecm}\bar{\xi}\hspace{-0.06truecm}-\hspace{-0.06truecm}u\hspace{-0.06truecm}-\hspace{-0.06truecm}\eta)},\no\\[6pt]
&& k^+_x(u)=\frac{\s(\hspace{-0.06truecm}-\hspace{-0.06truecm}2u\hspace{-0.06truecm}-\hspace{-0.06truecm}2\eta)
   \s_{(1,0)}(\l_1\hspace{-0.06truecm}+\hspace{-0.06truecm}\l_2\hspace{-0.06truecm}+\hspace{-0.06truecm}\eta
   \hspace{-0.06truecm}-\hspace{-0.06truecm}\frac{1}{2})\s_{(1,0)}(\l_1\hspace{-0.06truecm}+\hspace{-0.06truecm}\bar{\xi})
   \s_{(1,0)}(\l_2\hspace{-0.06truecm}+\hspace{-0.06truecm}\bar{\xi})}
   {2\s_{(1,0)}(\hspace{-0.06truecm}-\hspace{-0.06truecm}u\hspace{-0.06truecm}-\hspace{-0.06truecm}\eta)
   \s(u\hspace{-0.06truecm}+\hspace{-0.06truecm}\eta\hspace{-0.06truecm}+\hspace{-0.06truecm}\l_1\hspace{-0.06truecm}+\hspace{-0.06truecm}\l_2
   \hspace{-0.06truecm}-\hspace{-0.06truecm}\frac{1}{2})\s(\l_1\hspace{-0.06truecm}+\hspace{-0.06truecm}\bar{\xi}
   \hspace{-0.06truecm}-\hspace{-0.06truecm}u\hspace{-0.06truecm}-\hspace{-0.06truecm}\eta)
   \s(\l_2\hspace{-0.06truecm}+\hspace{-0.06truecm}\bar{\xi}\hspace{-0.06truecm}-\hspace{-0.06truecm}u\hspace{-0.06truecm}-\hspace{-0.06truecm}\eta)},\no\\[6pt]
&& k^+_y(u)=\frac{i\,\s(-\hspace{-0.06truecm}2u\hspace{-0.06truecm}-\hspace{-0.06truecm}2\eta)
   \s_{(1,1)}(\l_1\hspace{-0.06truecm}+\hspace{-0.06truecm}\l_2\hspace{-0.06truecm}+\hspace{-0.06truecm}\eta\hspace{-0.06truecm}-\hspace{-0.06truecm}\frac{1}{2})
   \s_{(1,1)}(\l_1\hspace{-0.06truecm}+\hspace{-0.06truecm}\bar{\xi})\s_{(1,1)}(\l_2\hspace{-0.06truecm}+\hspace{-0.06truecm}\bar{\xi})}
   {2\s_{(1,1)}(\hspace{-0.06truecm}-\hspace{-0.06truecm}u\hspace{-0.06truecm}-\hspace{-0.06truecm}\eta)
   \s(u\hspace{-0.06truecm}+\hspace{-0.06truecm}\eta\hspace{-0.06truecm}+\hspace{-0.06truecm}\l_1\hspace{-0.06truecm}+\hspace{-0.06truecm}
   \l_2\hspace{-0.06truecm}-\hspace{-0.06truecm}\frac{1}{2})
   \s(\l_1\hspace{-0.06truecm}+\hspace{-0.06truecm}\bar{\xi}\hspace{-0.06truecm}-\hspace{-0.06truecm}u\hspace{-0.06truecm}-\hspace{-0.06truecm}\eta)
   \s(\l_2\hspace{-0.06truecm}+\hspace{-0.06truecm}\bar{\xi}\hspace{-0.06truecm}-\hspace{-0.06truecm}u\hspace{-0.06truecm}-\hspace{-0.06truecm}\eta)},\no\\[6pt]
&& k^+_z(u)=\frac{\s(\hspace{-0.06truecm}-\hspace{-0.06truecm}2u\hspace{-0.06truecm}-\hspace{-0.06truecm}2\eta)
   \s_{(0,1)}(\l_1\hspace{-0.06truecm}+\hspace{-0.06truecm}\l_2\hspace{-0.06truecm}+\hspace{-0.06truecm}\eta\hspace{-0.06truecm}-\hspace{-0.06truecm}\frac{1}{2})
   \s_{(0,1)}(\l_1\hspace{-0.06truecm}+\hspace{-0.06truecm}\bar{\xi})\s_{(0,1)}(\l_2\hspace{-0.06truecm}+\hspace{-0.06truecm}\bar{\xi})}
   {2\s_{(0,1)}(\hspace{-0.06truecm}-\hspace{-0.06truecm}u\hspace{-0.06truecm}-\hspace{-0.06truecm}\eta)
   \s(u\hspace{-0.06truecm}+\hspace{-0.06truecm}\eta\hspace{-0.06truecm}+\hspace{-0.06truecm}\l_1\hspace{-0.06truecm}+\hspace{-0.06truecm}\l_2
   \hspace{-0.06truecm}-\hspace{-0.06truecm}\frac{1}{2})
   \s(\l_1\hspace{-0.06truecm}+\hspace{-0.06truecm}\bar{\xi}\hspace{-0.06truecm}-\hspace{-0.06truecm}u\hspace{-0.06truecm}-\hspace{-0.06truecm}\eta)
   \s(\l_2\hspace{-0.06truecm}+\hspace{-0.06truecm}\bar{\xi}\hspace{-0.06truecm}-\hspace{-0.06truecm}u\hspace{-0.06truecm}-\hspace{-0.06truecm}\eta)}.\label{K-matrix-6}
\eea The K-matrices $K^{\mp}(u)$ depend on four free boundary parameters
$\{\l_1,\,\l_2,\,\xi,\,\bar{\xi}\}$. It is
very convenient to introduce a vector $\l\in V$ associated with
the boundary parameters $\{\l_i\}$, \bea
 \l=\sum_{k=1}^2\l_k\e_k. \label{boundary-vector}
\eea


\subsection{Vertex-face correspondence}

Let us briefly review the face-type R-matrix associated with the
six-vertex model.

Set \bea \hat{\imath}=\e_i-\overline{\e},~~\overline{\e}=
\frac{1}{2}\sum_{k=1}^{2}\e_k, \quad i=1,2,\qquad {\rm then}\,
\sum_{i=1}^2\hat{\imath}=0. \label{fundmental-vector} \eea Let
$\h$ be the Cartan subalgebra of $A_{1}$ and $\h^{*}$ be its dual.
A finite dimensional diagonalizable  $\h$-module is a complex
finite dimensional vector space $W$ with a weight decomposition
$W=\oplus_{\mu\in \h^*}W[\mu]$, so that $\h$ acts on $W[\mu]$ by
$x\,v=\mu(x)\,v$, $(x\in \h,\,v\in\,W[\mu])$. For example, the
non-zero weight spaces of the fundamental representation
$V_{\L_1}=\Cb^2=V$ are
\bea
 W[\hat{\imath}]=\Cb \e_i,~i=1,2.\label{Weight}
\eea

For a generic $m\in V$, define \bea m_i=\langle m,\e_i\rangle,
~~m_{ij}=m_i-m_j=\langle m,\e_i-\e_j\rangle,~~i,j=1,2.
\label{Def1}\eea Let $R(u,m)\in {\rm End}(V\otimes V)$ be the
R-matrix of the eight-vertex SOS model \cite{Bax82} given by
\bea
R(u;m)\hspace{-0.1cm}=\hspace{-0.1cm}
\sum_{i=1}^{2}R(u;m)^{ii}_{ii}E_{ii}\hspace{-0.1cm}\otimes\hspace{-0.1cm}
E_{ii}\hspace{-0.1cm}+\hspace{-0.1cm}\sum_{i\ne
j}^2\lt\{R(u;m)^{ij}_{ij}E_{ii}\hspace{-0.1cm}\otimes\hspace{-0.1cm}
E_{jj}\hspace{-0.1cm}+\hspace{-0.1cm}
R(u;m)^{ji}_{ij}E_{ji}\hspace{-0.1cm}\otimes\hspace{-0.1cm}
E_{ij}\rt\}, \label{R-matrix} \eea where $E_{ij}$ is the matrix
with elements $(E_{ij})^l_k=\d_{jk}\d_{il}$. The coefficient
functions are \bea
 &&R(u;m)^{ii}_{ii}=1,~~
   R(u;m)^{ij}_{ij}=\frac{\s(u)\s(m_{ij}-\eta)}
   {\s(u+\eta)\s(m_{ij})},~~i\neq j,\label{Elements1}\\
 && R(u;m)^{ji}_{ij}=\frac{\s(\eta)\s(u+m_{ij})}
    {\s(u+\eta)\s(m_{ij})},~~i\neq j,\label{Elements2}
\eea  and $m_{ij}$ is defined in (\ref{Def1}). The R-matrix
satisfies the dynamical (modified) quantum Yang-Baxter equation
(or the star-triangle relation) \cite{Bax82}
\begin{eqnarray}
&&R_{1,2}(u_1-u_2;m-\eta h^{(3)})R_{1,3}(u_1-u_3;m)
R_{2,3}(u_2-u_3;m-\eta h^{(1)})\no\\
&&\qquad =R_{2,3}(u_2-u_3;m)R_{1,3}(u_1-u_3;m-\eta
h^{(2)})R_{1,2}(u_1-u_2;m).\label{MYBE}
\end{eqnarray}
Here we have adopted
\bea R_{1,2}(u,m-\eta h^{(3)})\,v_1\otimes
v_2 \otimes v_3=\lt(R(u,m-\eta\mu)\otimes {\rm id }\rt)v_1\otimes
v_2 \otimes v_3,\quad {\rm if}\, v_3\in W[\mu]. \label{Action}
\eea Moreover, one may check that the R-matrix satisfies  the weight
conservation condition, \bea
  \lt[h^{(1)}+h^{(2)},\,R_{1,2}(u;m)\rt]=0,\label{Conservation}
\eea the unitary condition, \bea
 R_{1,2}(u;m)\,R_{2,1}(-u;m)={\rm id}\otimes {\rm
 id},\label{Unitary}
\eea and the crossing relation \bea
 R(u;m)^{kl}_{ij}=\varepsilon_{l}\,\varepsilon_{j}
   \frac{\s(u)\s((m-\eta\hat{\imath})_{21})}
   {\s(u+\eta)\s(m_{21})}R(-u-\eta;m-\eta\hat{\imath})
   ^{\bar{j}\,k}_{\bar{l}\,i},\label{Crossing}
\eea where
\bea \varepsilon_{1}=1,\,\varepsilon_{2}=-1,\quad {\rm
and}\,\, \bar{1}=2,\,\bar{2}=1.\label{Parity} \eea

Let us introduce two intertwiners which are
$2$-component  column vectors $\phi_{m,m-\eta\hat{\jmath}}(u)$
labelled by $\hat{1},\,\hat{2}$. The $k$-th element of
$\phi_{m,m-\eta\hat{\jmath}}(u)$ is given by \bea
\phi^{(k)}_{m,m-\eta\hat{\jmath}}(u)=\theta^{(k)}(u+2m_j),\label{Intvect}\eea
where the functions $\theta^{(j)}(u)$ are given in (\ref{Function-j}).
Explicitly,
\bea \phi_{m,m-\eta\hat{1}}(u)=
\lt(\begin{array}{c}\theta^{(1)}(u+2m_1)\\[6pt]\theta^{(2)}(u+2m_1)\end{array}\rt),\qquad
\phi_{m,m-\eta\hat{2}}(u)=
\lt(\begin{array}{c}\theta^{(1)}(u+2m_2)\\[6pt]\theta^{(2)}(u+2m_2)\end{array}\rt).\eea
One can prove the following identity \cite{Fan98}
\bea
 {\rm det}\lt|\begin{array}{cc}\theta^{(1)}(u+2m_1)&\theta^{(1)}(u+2m_2)\\[6pt]
   \theta^{(2)}(u+2m_1)&\theta^{(2)}(u+2m_2)\end{array}\rt|=C(\tau)\,\s(u+m_1+m_2-\frac{1}{2})
   \,\s(m_{12}),\no
\eea where $C(\tau)$ is non-vanishing constant which only depends on  $\tau$. This implies that
the two intertwiner vectors
$\phi_{m,m-\eta\hat{\imath}}(u)$ are linearly {\it independent}
for a generic $m\in V$.

 Using the intertwiner vectors, one can derive the following face-vertex
correspondence relation \cite{Bax82}\bea &&\R_{1,2}(u_1-u_2)
\phi^1_{m,m-\eta\hat{\imath}}(u_1)
\phi^2_{m-\eta\hat{\imath},m-\eta(\hat{\imath}+\hat{\jmath})}(u_2)
\no\\&&~~~~~~= \sum_{k,l}R(u_1-u_2;m)^{kl}_{ij}
\phi^1_{m-\eta\hat{l},m-\eta(\hat{l}+\hat{k})}(u_1)
\phi^2_{m,m-\eta\hat{l}}(u_2). \label{Face-vertex} \eea  Then the
QYBE (\ref{QYBE}) of the vertex-type R-matrix $\R(u)$ is equivalent
to the dynamical Yang-Baxter equation (\ref{MYBE}) of the SOS
R-matrix $R(u,m)$. For a generic $m$, we can introduce other types
of intertwiners $\bar{\phi},~\tilde{\phi}$ which  are both row
vectors and satisfy the following conditions, \bea
  &&\bar{\phi}_{m,m-\eta\hat{\mu}}(u)
     \,\phi_{m,m-\eta\hat{\nu}}(u)=\d_{\mu\nu},\quad
     \tilde{\phi}_{m+\eta\hat{\mu},m}(u)
     \,\phi_{m+\eta\hat{\nu},m}(u)=\d_{\mu\nu},\label{Int2}\eea
{}from which one  can derive the relations,
\begin{eqnarray}
&&\sum_{\mu=1}^2\phi_{m,m-\eta\hat{\mu}}(u)\,
 \bar{\phi}_{m,m-\eta\hat{\mu}}(u)={\rm id},\label{Int3}\\
&&\sum_{\mu=1}^2\phi_{m+\eta\hat{\mu},m}(u)\,
 \tilde{\phi}_{m+\eta\hat{\mu},m}(u)={\rm id}.\label{Int4}
\end{eqnarray}
With the help of (\ref{Face-vertex})-(\ref{Int4}), we obtain,
\begin{eqnarray}
 &&\tilde{\phi}^1_{m+\eta\hat{k},m}(u_1)\,\R_{1,2}(u_1-u_2)
 \phi^2_{m+\eta\hat{\jmath},m}(u_2)\no\\
 &&\qquad\quad= \sum_{i,l}R(u_1-u_2;m)^{kl}_{ij}\,
 \tilde{\phi}^1_{m+\eta(\hat{\imath}+\hat{\jmath}),m+\eta\hat{\jmath}}(u_1)
 \phi^2_{m+\eta(\hat{k}+\hat{l}),m+\eta\hat{k}}(u_2),\label{Face-vertex1}\\
 &&\tilde{\phi}^1_{m+\eta\hat{k},m}(u_1)
 \tilde{\phi}^2_{m+\eta(\hat{k}+\hat{l}),m+\eta\hat{k}}(u_2)\,
 \R_{1,2}(u_1-u_2)\no\\
 &&\qquad\quad= \sum_{i,j}R(u_1-u_2;m)^{kl}_{ij}\,
 \tilde{\phi}^1_{m+\eta(\hat{\imath}+\hat{\jmath}),m+\eta\hat{\jmath}}(u_1)
 \tilde{\phi}^2_{m+\eta\hat{\jmath},m}(u_2),\label{Face-vertex2}\\
 &&\bar{\phi}^2_{m,m-\eta\hat{l}}(u_2)\,\R_{1,2}(u_1-u_2)
   \phi^1_{m,m-\eta\hat{\imath}}(u_1)\no\\
 &&\qquad\quad= \sum_{k,j}R(u_1-u_2;m)^{kl}_{ij}\,
 \phi^1_{m-\eta\hat{l},m-\eta(\hat{k}+\hat{l})}(u_1)
 \bar{\phi}^2_{m-\eta\hat{\imath},m-\eta(\hat{\imath}+\hat{\jmath})}(u_2),\label{Face-vertex3}\\
 &&\bar{\phi}^1_{m-\eta\hat{l},m-\eta(\hat{k}+\hat{l})}(u_1)
 \bar{\phi}^2_{m,m-\eta\hat{l}}(u_2)\,\R_{12}(u_1-u_2)\no\\
 &&\qquad\quad= \sum_{i,j}R(u_1-u_2;m)^{kl}_{ij}\,
 \bar{\phi}^1_{m,m-\eta\hat{\imath}}(u_1)
 \bar{\phi}^2_{m-\eta\hat{\imath},m-\eta(\hat{\imath}
 +\hat{\jmath})}(u_2).\label{Face-vertex4}
\end{eqnarray}

In addition to the Riemann identity (\ref{identity}), the $\s$-function enjoys the
following properties:
\bea
 &&\s(2u)=\frac{2\s(u)\,\s_{(0,1)}(u)\,\s_{(1,0)}(u)\,\s_{(1,1)}(u)}
     {\s_{(0,1)}(0)\,\s_{(1,0)}(0)\,\s_{(1,1)}(0)},\\
 &&\s(u+1)=-\s(u),\quad\quad \s(u+\tau)=e^{-2i\pi(u+\frac{1}{2}+\frac{\tau}{2})}\s(u),
\eea where the functions $\s_{a}(u)$ are given by (\ref{sigma-function}).
Using the above identities and  the method in \cite{Fan98}, after tedious calculations,
we can show that the K-matrices $K^{\pm}(u)$ given by
(\ref{K-matrix}) and (\ref{DK-matrix}) can be expressed in terms
of the intertwiners and {\it diagonal\/} matrices $\K(\l|u)$ and
$\tilde{\K}(\l|u)$ as follows \bea &&K^-(u)^s_t=
\sum_{i,j}\phi^{(s)}_{\l-\eta(\hat{\imath}-\hat{\jmath}),
~\l-\eta\hat{\imath}}(u)
\K(\l|u)^j_i\bar{\phi}^{(t)}_{\l,~\l-\eta\hat{\imath}}(-u),\label{K-F-1}\\
&&K^+(u)^s_t= \sum_{i,j}
\phi^{(s)}_{\l,~\l-\eta\hat{\jmath}}(-u)\tilde{\K}(\l|u)^j_i
\tilde{\phi}^{(t)}_{\l-\eta(\hat{\jmath}-\hat{\imath}),
~\l-\eta\hat{\jmath}}(u).\label{K-F-2}\eea Here the two {\it
diagonal\/} matrices $\K(\l|u)$ and $\tilde{\K}(\l|u)$ are given
by \bea
&&\K(\l|u)\equiv{\rm Diag}(k(\l|u)_1,\,k(\l|u)_2)={\rm
Diag}(\frac{\s(\l_1+\xi-u)}{\s(\l_1+\xi+u)},\,
\frac{\s(\l_2+\xi-u)}{\s(\l_2+\xi+u)}),\label{K-F-3}\\
&&\tilde{\K}(\l|u)\equiv{\rm
Diag}(\tilde{k}(\l|u)_1,\,\tilde{k}(\l|u)_2)\no\\
&&~~~~~~~~~={\rm
Diag}(\frac{\s(\l_{12}\hspace{-0.1cm}-\hspace{-0.1cm}
\eta)\s(\l_1\hspace{-0.1cm}+\hspace{-0.1cm}\bar{\xi}+\hspace{-0.1cm}u
\hspace{-0.1cm}+\hspace{-0.1cm}\eta)}
{\s(\l_{12})\s(\l_1+\bar{\xi}-u-\eta)},\,
\frac{\s(\l_{12}\hspace{-0.1cm}+\hspace{-0.1cm}
\eta)\s(\l_2\hspace{-0.1cm}+\hspace{-0.1cm}\bar{\xi}\hspace{-0.1cm}
+\hspace{-0.1cm}u\hspace{-0.1cm}+\hspace{-0.1cm}\eta)}
{\s(\l_{12})\s(\l_2+\bar{\xi}-u-\eta)}).\label{K-F-4} \eea
Although the vertex type K-matrices $K^{\pm}(u)$ given by
(\ref{K-matrix}) and (\ref{DK-matrix}) are generally non-diagonal,
after the face-vertex transformations (\ref{K-F-1}) and
(\ref{K-F-2}), the face type counterparts $\K(\l|u)$ and
$\tilde{\K}(\l|u)$  become {\it simultaneously\/} diagonal. This
fact enabled the authors in \cite{Yan04-1,Yan07} to diagonalize the transfer matrices $\t(u)$ (\ref{trans})
by applying the generalized algebraic Bethe
ansatz method developed in \cite{Yan04}.
.

\subsection{Two sets of eigenstates}

In order to construct the Bethe states of the open XYZ model
with non-diagonal boundary terms specified by the K-matrices
(\ref{K-matrix-2-1}) and (\ref{K-matrix-6}), we need to introduce
the new double-row monodromy matrices $\T^{\pm}(m|u)$ \cite{Yan04,Yan10,Fen10}:
\bea
 \T^-(m|u)^{\nu}_{\mu}
     &=&\tilde{\phi}^{0}_{m-\eta(\hat{\mu}-\hat{\nu}),
        m-\eta\hat{\mu}}(u)~\mathbb{T}_0(u)\phi^{0}_{m,
        m-\eta\hat{\mu}}(-u),\label{Mon-F}\\
\T^+(m|u)^j_i
     &=&\prod_{k\neq j}\frac{\s(m_{jk})}{\s(m_{jk}-\eta)}
        \,\phi^{t_0}_{m-\eta(\hat{\jmath}-\hat{\imath}),m-\eta\hat{\jmath}}(u)
        \lt(\mathbb{T}^+(u)\rt)^{t_0}\bar{\phi}^{t_0}_{m,m-\eta\hat{\jmath}}(-u),
        \label{Mon-F-1}
\eea where $t_0$ denotes transposition in the $0$-th space (i.e.
auxiliary space) and $\mathbb{T}^+(u)$ is given by
\bea
  \lt(\mathbb{T}^+(u)\rt)^{t_0}&=&T^{t_0}(u)\lt(K^+(u)\rt)^{t_0}\hat{T}^{t_0}(u).
      \label{Mon-V-1}
\eea
These double-row monodromy matrices, in the face picture, can be
expressed in terms of the face type R-matrix $R(u;m)$
(\ref{R-matrix}) and  K-matrices  $\K(\l|u)$ (\ref{K-F-3}) and $\tilde{\K}(\l|u)$
(\ref{K-F-4}) (for the details see section 3).

So far  only two sets of Bethe states ( i.e. eigenstates) of the transfer matrix
for the models with non-diagonal boundary terms  have been found \cite{Fan96,Yan07,Fen10}.
These two sets of states are
\bea
&&|\{v^{(1)}_i\}\rangle^{(I)}=
     \T^+(\l+2\eta\hat{1}|v^{(1)}_1)^1_2\cdots
   \T^+(\l+2M\eta\hat{1}|v^{(1)}_M)^1_2|\O^{(I)}(\l)\rangle,
   \label{Bethe-state-1}\\
&&|\{v^{(2)}_i\}\rangle^{(II)} =
   \T^-(\l-2\eta\hat{2}|v^{(2)}_1)^2_1
   \cdots
   \T^-(\l\hspace{-0.04truecm}-\hspace{-0.04truecm}2M\eta\hat{2}|v^{(2)}_M)^2_1|\O^{(II)}(\l)\rangle,
   \label{Bethe-state-2}
\eea where the vector $\l$ is related to the boundary parameters
(\ref{boundary-vector}). The associated reference states
$|\O^{(I)}(\l)\rangle$ and $|\O^{(II)}(\l)\rangle$ are \bea
\hspace{-1.2truecm}|\O^{(I)}(\l)\rangle
   &=&\phi^1_{\l+N\eta\hat{1},\l+(N-1)\eta\hat{1}}(z_1)
      \phi^2_{\l+(N-1)\eta\hat{1},\l+(N-2)\eta\hat{1}}(z_{2})\cdots
      \phi^N_{\l+\eta\hat{1},\l}(z_N),\label{Vac-1}\\
\hspace{-1.2truecm} |\O^{(II)}(\l)\rangle&=&
\phi^1_{\l,\l-\eta\hat{2}}(z_1)
\phi^{2}_{\l-\eta\hat{2},\l-2\eta\hat{2}}(z_{2})\cdots
\phi^N_{\l-(N-1)\eta\hat{2},\l-N\eta\hat{2}}(z_N).\label{Vac-2}
\eea It is remarked that   $\phi^k={\rm id}\otimes {\rm
id}\cdots\otimes \stackrel{k-th}{\phi}\otimes {\rm id}\cdots$.

If the parameters $\{v^{(1)}_k\}$ satisfy the first set of  Bethe
ansatz equations given by
\bea &&\hspace{-0.1cm}\frac
{\s(\l_2+\xi+v^{(1)}_{\a})\s(\l_2+\bar\xi-v^{(1)}_{\a})
\s(\l_1+\bar\xi+v^{(1)}_{\a})\s(\l_1+\xi-v^{(1)}_{\a})}
{\s(\l_2\hspace{-0.1cm}+\hspace{-0.1cm}
\bar\xi\hspace{-0.1cm}+\hspace{-0.1cm}v^{(1)}_{\a}
\hspace{-0.1cm}+\hspace{-0.1cm}\eta)
\s(\l_2\hspace{-0.1cm}+\hspace{-0.1cm}\xi\hspace{-0.1cm}-\hspace{-0.1cm}v^{(1)}_{\a}
\hspace{-0.1cm}-\hspace{-0.1cm}\eta)
\s(\l_1\hspace{-0.1cm}+\hspace{-0.1cm}\xi\hspace{-0.1cm}+\hspace{-0.1cm}
v^{(1)}_{\a}\hspace{-0.1cm}+\hspace{-0.1cm}\eta)
\s(\l_1\hspace{-0.1cm}+\hspace{-0.1cm}\bar\xi\hspace{-0.1cm}-\hspace{-0.1cm}v^{(1)}_{\a}
\hspace{-0.1cm}-\hspace{-0.1cm}\eta)}\no\\
&&~~~~~~=\prod_{k\neq
\a}^M\frac{\s(v^{(1)}_{\a}+v^{(1)}_k+2\eta)\s(v^{(1)}_{\a}-v^{(1)}_k+\eta)}
{\s(v^{(1)}_{\a}+v^{(1)}_k)\s(v^{(1)}_{\a}-v^{(1)}_k-\eta)}\no\\
&&~~~~~~~~~~\times\prod_{k=1}^{2M}\frac{\s(v^{(1)}_{\a}+z_k)\s(v^{(1)}_{\a}-z_k)}
{\s(v^{(1)}_{\a}+z_k+\eta)\s(v^{(1)}_{\a}-z_k+\eta)},~~\a=1,\cdots,M,
\label{BA-D-1}\eea the Bethe state
$|v^{(1)}_1,\cdots,v^{(1)}_M\rangle^{(1)}$ becomes the eigenstate
of the transfer matrix with eigenvalue $\L^{(1)}(u)$  given by
\cite{Fen10}
\bea
&&\L^{(1)}(u)=\frac{\s(\l_2+\bar\xi-u)\s(\l_1+\bar\xi+u)\s(\l_1+\xi-u)\s(2u+2\eta)}
{\s(\l_2+\bar\xi-u-\eta)\s(\l_1+\bar\xi-u-\eta)\s(\l_1+\xi+u)\s(2u+\eta)}\no\\
&&~~~~~~~~~~~~~~~~~~\times\prod_{k=1}^M\frac{\s(u+v^{(1)}_k)\s(u-v^{(1)}_k-\eta)}
{\s(u+v^{(1)}_k+\eta)\s(u-v^{(1)}_k)}\no\\
&&~~~~~~+\frac{\s(\l_2+\bar\xi+u+\eta)\s(\l_1+\xi+u+\eta)\s(\l_2+\xi-u-\eta)\s(2u)}
{\s(\l_2+\bar\xi-u-\eta)\s(\l_1+\xi+u)\s(\l_2+\xi+u)\s(2u+\eta)}\no\\
&&~~~~~~~~~~~~~~~~~~\times\prod_{k=1}^M\frac{\s(u+v^{(1)}_k+2\eta)\s(u-v^{(1)}_k+\eta)}
{\s(u+v^{(1)}_k+\eta)\s(u-v^{(1)}_k)}\no\\
&&~~~~~~~~~~~~~~~~~~\times\prod_{k=1}^{2M}\frac{\s(u+z_k)\s(u-z_k)}
{\s(u+z_k+\eta)\s(u-z_k+\eta)}.\label{Eigenfuction-D-1}
 \eea

\noindent If the parameters $\{v^{(2)}_k\}$ satisfy the second
Bethe Ansatz equations
\bea
&&\hspace{-0.1cm}\frac
  {\s(\l_1+\xi+v^{(2)}_{\a})\s(\l_1+\bar\xi-v^{(2)}_{\a})
  \s(\l_2+\bar\xi+v^{(2)}_{\a})\s(\l_2+\xi-v^{(2)}_{\a})}
  {\s(\l_1\hspace{-0.1cm}+\hspace{-0.1cm}
  \bar\xi\hspace{-0.1cm}+\hspace{-0.1cm}v^{(2)}_{\a}
  \hspace{-0.1cm}+\hspace{-0.1cm}\eta)
  \s(\l_1\hspace{-0.1cm}+\hspace{-0.1cm}\xi\hspace{-0.1cm}-\hspace{-0.1cm}v^{(2)}_{\a}
  \hspace{-0.1cm}-\hspace{-0.1cm}\eta)
  \s(\l_2\hspace{-0.1cm}+\hspace{-0.1cm}\xi\hspace{-0.1cm}+\hspace{-0.1cm}
  v^{(2)}_{\a}\hspace{-0.1cm}+\hspace{-0.1cm}\eta)
  \s(\l_2\hspace{-0.1cm}+\hspace{-0.1cm}\bar\xi\hspace{-0.1cm}-\hspace{-0.1cm}v^{(2)}_{\a}
  \hspace{-0.1cm}-\hspace{-0.1cm}\eta)}\no\\
&&~~~~~~=\prod_{k\neq
  \a}^M\frac{\s(v^{(2)}_{\a}+v^{(2)}_k+2\eta)\s(v^{(2)}_{\a}-v^{(2)}_k+\eta)}
  {\s(v^{(2)}_{\a}+v^{(2)}_k)\s(v^{(2)}_{\a}-v^{(2)}_k-\eta)}\no\\
&&~~~~~~~~~~\times\prod_{k=1}^{2M}\frac{\s(v^{(2)}_{\a}+z_k)\s(v^{(2)}_{\a}-z_k)}
  {\s(v^{(2)}_{\a}+z_k+\eta)\s(v^{(2)}_{\a}-z_k+\eta)},~~\a=1,\cdots,M,
  \label{BA-D-2}
\eea the Bethe states $|v^{(2)}_1,\cdots,v^{(2)}_M\rangle^{(II)}$
yield the second set of the eigenstates of the transfer matrix
with the eigenvalues \cite{Fan96,Yan07},
\bea
&&\L^{(2)}(u)=\frac{\s(2u+2\eta)\s(\l_1+\bar\xi-u)\s(\l_2+\bar\xi+u)\s(\l_2+\xi-u)}
{\s(2u+\eta)\s(\l_1+\bar\xi-u-\eta)\s(\l_2+\bar\xi-u-\eta)\s(\l_2+\xi+u)}\no\\
&&~~~~~~~~~~~~~~~~~~\times\prod_{k=1}^M\frac{\s(u+v^{(2)}_k)\s(u-v^{(2)}_k-\eta)}
{\s(u+v^{(2)}_k+\eta)\s(u-v^{(2)}_k)}\no\\
&&~~~~~~+\frac{\s(2u)\s(\l_1+\bar\xi+u+\eta)
\s(\l_2+\xi+u+\eta)\s(\l_1+\xi-u-\eta)}
{\s(2u+\eta)\s(\l_1+\bar\xi-u-\eta)\s(\l_2+\xi+u)\s(\l_1+\xi+u)}\no\\
&&~~~~~~~~~~~~~~~~~~\times\prod_{k=1}^M\frac{\s(u+v^{(2)}_k+2\eta)\s(u-v^{(2)}_k+\eta)}
{\s(u+v^{(2)}_k+\eta)\s(u-v^{(2)}_k)}\no\\
&&~~~~~~~~~~~~~~~~~~\times\prod_{k=1}^{2M}\frac{\s(u+z_k)\s(u-z_k)}
{\s(u+z_k+\eta)\s(u-z_k+\eta)}.\label{Eigenfuction-D-2}
 \eea


\section{ $\T^{\pm}(m|u)$ in the face picture}
\label{T} \setcounter{equation}{0}

The K-matrices $K^{\pm}(u)$ given by (\ref{K-matrix}) and
(\ref{DK-matrix}) are generally non-diagonal (in the vertex
picture), after the face-vertex transformations (\ref{K-F-1}) and
(\ref{K-F-2}), the face type counterparts $\K(\l|u)$ and
$\tilde{\K}(\l|u)$ given by (\ref{K-F-3}) and (\ref{K-F-4}) {\it
simultaneously\/} become diagonal. This fact suggests that it
would be much simpler if one performs all calculations in the face
picture \cite{Yan11}.

Let us introduce the face type one-row monodromy matrix (c.f
(\ref{Mon-V})) \bea
 T_{F}(l|u)&\equiv &T^{F}_{0,1\ldots N}(l|u)\no\\
 &=&R_{0,N}(u-z_N;l-\eta\sum_{i=1}^{N-1}h^{(i)})\ldots
    R_{0,2}(u-z_2;l-\eta h^{(1)})R_{0,1}(u-z_1;l),\no\\
 &=&\lt(\begin{array}{ll}T_F(l|u)^1_1&T_F(l|u)^1_2\\T_F(l|u)^2_1&
   T_F(l|u)^2_2\end{array}\rt)
    \label{Monodromy-face-1}
\eea where $l$ is a generic vector in $V$. The monodromy matrix
satisfies the face type quadratic exchange relation
\cite{Fel96,Hou03}. Applying $T_F(l|u)^i_j$ to an arbitrary vector
$|i_1,\ldots,i_N\rangle$ in the N-tensor product space $V^{\otimes
N}$ given by \bea
   |i_1,\ldots,i_N\rangle=\e^1_{i_1}\ldots
   \e^N_{i_N},\label{Vector-V}
\eea we have \bea
 T_F(l|u)^i_j|i_1,\ldots,i_N\rangle&\equiv&
    T_F(m;l|u)^i_j|i_1,\ldots,i_N\rangle\no\\
 &=&\sum_{\a_{N-1}\ldots\a_1}\sum_{i'_N\ldots i'_1}
 R(u-z_N;l-\eta\sum_{k=1}^{N-1}\hat{\imath}'_k)
   ^{i\,\,\,\,\,\,\,\,\,\,\,\,\,\,i'_N}_{\a_{N-1}\,i_N}\ldots\no\\
 &&\quad\quad\times R(u-z_2;l-\eta\hat{\imath}'_1)^{\a_2\,i'_2}_{\a_1\,\,i_2}
 R(u-z_1;l)^{\a_1\,i'_1}_{j\,\,\,\,i_1}
   \,\,|i'_1,\ldots,i'_N\rangle,\label{Monodromy-face-2}
\eea where $m=l-\eta\sum_{k=1}^N\hat{\imath}_k$. We shall express
the double-row monodromy matrices $\T^{\pm}$ given by
(\ref{Mon-F}) and (\ref{Mon-F-1}) in terms of the above face-type
one-row monodromy matrix.

Associated with the vertex type monodromy matrices $T(u)$
(\ref{Mon-V}) and $\hat{T}(u)$ (\ref{Mon-V-0}), we introduce the
following operators \bea
 T(m,l|u)^j_{\mu}&=&\tilde{\phi}^0_{m+\eta\hat{\jmath},m}(u)\,T_0(u)\,
    \phi^0_{l+\eta\hat{\mu},l}(u),\\
 S(m,l|u)^{\mu}_{i}&=&\bar{\phi}^0_{l,l-\eta\hat{\mu}}(-u)\,\hat{T}_0(u)\,
    \phi^0_{m,m-\eta\hat{\imath}}(-u).
\eea Moreover, for the case of
$m=l-\eta\sum_{k=1}^N\hat{\imath}_k$, we introduce a generic state
in the quantum space from  the intertwiner vector (\ref{Intvect})
\bea
 |i_1,\ldots,i_N\rangle^{m}_{l}=
     \phi^1_{l,l-\eta\hat{\imath}_1}(z_1)
     \phi^2_{l-\eta\hat{\imath}_1,l-\eta(\hat{\imath}_1+\hat{\imath}_2)}(z_2)\ldots
     \phi^N_{l-\eta\sum_{k=1}^{N-1}\hat{\imath}_k,l-\eta\sum_{k=1}^{N}\hat{\imath}_k}(z_N).
\eea We can evaluate the action of the operator $T(m,l|u)$ on the
state $|i_1,\ldots,i_N\rangle^{m}_{l}$ from the face-vertex
correspondence relation (\ref{Face-vertex}) \bea
 &&T(m,l|u)^j_{\mu}|i_1,\ldots,i_N\rangle^{m}_{l}=
    \tilde{\phi}^0_{m+\eta\hat{\jmath},m}(u)\,T_0(u)\,
    \phi^0_{l+\eta\hat{\mu},l}(u)|i_1,\ldots,i_N\rangle^{m}_{l}
    \no\\
 &&\quad\quad=\tilde{\phi}^0_{m+\eta\hat{\jmath},m}(u)
    \R_{0,N}(u-z_N)\ldots\R_{0,1}(u-z_1)\phi^0_{l+\eta\hat{\mu},l}(u)
    \phi^1_{l,l-\eta\hat{\imath}_1}(z_1)\ldots\no\\
 &&\quad\quad=\sum_{\a_1,i'_1}R(u-z_1;l+\eta\hat{\mu})^{\a_1i'_1}_{\mu\,\, i_1}
   \phi^1_{l+\eta\hat{\mu},l+\eta\hat{\mu}-\eta\hat{\imath}'_1}(z_1)
   \tilde{\phi}^0_{m+\eta\hat{\jmath},m}(u)
    \R_{0,N}(u-z_N)\ldots\no\\
 &&\quad\quad\quad\quad \times
   \R_{0,2}(u-z_2)\phi^0_{l+\eta\hat{\mu}-\eta\hat{\imath}'_1,l-\eta\hat{\imath}_1}(u)
    \phi^2_{l-\eta\hat{\imath}_1,l-\eta(\hat{\imath}_1+\hat{\imath}_2)}(z_2)\ldots\no\\
 &&\quad\quad\vdots\no\\
 &&\quad\quad=\sum_{\a_{1}\ldots\a_{N-1}}\sum_{i'_1\ldots i'_N}
    R(u-z_N;l+\eta\hat{\mu}-\eta\sum_{k=1}^{N-1}\hat{\imath}'_k)
    ^{j\,\,\,\,\,\,\,\,\,\,\,\,i'_N}_{\a_{N-1}i_N}\ldots\no\\
 &&\quad\quad\quad\quad\times R(u-z_1;l+\eta\hat{\mu})^{\a_1i'_1}_{\mu \,\,i_1}
    |i'_1,\ldots,i'_N\rangle^{l+\eta\hat{\mu}-\eta\sum_{k=1}^N\hat{\imath}'_k}_{l+\eta\hat{\mu}}.
    \label{T-1}
\eea  Here we have used the following property of the R-matrix
\bea
  R(u;m)^{i'j'}_{ij}=R(u;m\pm\eta(\hat{\imath}+\hat{\jmath}))^{i'j'}_{ij}
  =R(u;m\pm\eta(\hat{\imath}'+\hat{\jmath}'))^{i'j'}_{ij},\no
\eea and the weight conservation condition (\ref{Conservation}).
Comparing with (\ref{Monodromy-face-2}), we have the following
correspondence \bea
 T(m,l|u)^j_{\mu}|i_1,\ldots,i_N\rangle^{m}_{l}\,\longleftrightarrow\,
      T_F(m+\eta\hat{\mu};l+\eta\hat{\mu}|u)_{\mu}^j|i_1,\ldots,i_N\rangle,\label{crospendence-1}
\eea where vector $|i_1,\ldots,i_N\rangle$ is given by
(\ref{Vector-V}). Hereafter, we will use  $O_F$ to denote the face
version of operator $O$ in the face picture.

Noting that \bea
 \hat{T}_0(u)=\R_{1,0}(u+z_1)\ldots\R_{N,0}(u+z_N),\no
\eea we obtain the action of $S(m,l|u)^{\mu}_{i}$ on the state
$|i_1,\ldots,i_N\rangle^{m}_{l}$ \bea
 S(m,l|u)^{\mu}_{i}|i_1,\ldots,i_N\rangle^{m}_{l}
     \hspace{-0.22truecm}&=&\hspace{-0.42truecm}
     \sum_{\a_{1}\ldots\a_{N-1}}\sum_{i'_1\ldots i'_N}
     R(u+z_1;l)^{i'_1\,\mu}_{i_1\a_{N-1}}
     R(u+z_2;l-\eta\hat{\imath}_1)^{i'_2\,\a_{N-1}}_{i_2\a_{N-2}}\no\\
 &&\,\times
     \ldots R(u\hspace{-0.12truecm}+\hspace{-0.12truecm}z_N;
     l\hspace{-0.12truecm}-\hspace{-0.12truecm}\eta
     \sum_{k=1}^{N-1}\hat{\imath}_k)
     ^{i'_N\,\a_1}_{i_N\,i}|i'_1,\ldots,i'_N\rangle
     ^{l-\eta\hat{\mu}-\eta\sum_{k=1}^N\hat{\imath}'_k}_{l-\eta\hat{\mu}}.\no\\
\eea Then the crossing relation of the R-matrix (\ref{Crossing})
enables us to establish the following relation:
\bea
  S(m,l|u)^{\mu}_{i}=\varepsilon_{\bar{i}}\varepsilon_{\bar{\mu}}
   \frac{\s\lt(m_{21}\rt)}{\s\lt(l_{21}\rt)}
   \prod_{k=1}^N\frac{\s(u+z_k)}{\s(u+z_k+\eta)}
   T(m,l|-u-\eta)^{\bar{i}}_{\bar{\mu}},
   \label{Crossing-operator}
\eea where the parities are defined in (\ref{Parity}) and $m_{21}$
(or $l_{21}$) is defined in (\ref{Def1}).

Now we are in the position to express $\T^{\pm}$ (\ref{Mon-F}) and
(\ref{Mon-F-1}) in terms of $T(m,l)^i_j$ and $S(l,m)^i_j$. By
(\ref{Int3}) and (\ref{Int4}),  we have \bea
  \T^-(m|u)^j_i&=&\tilde{\phi}^{0}_{m-\eta(\hat{\imath}-\hat{\jmath}),
      m-\eta\hat{\imath}}(u)~\mathbb{T}(u)~\phi^{0}_{m,
      m-\eta\hat{\imath}}(-u)\no\\
  &=&\tilde{\phi}^{0}_{m-\eta(\hat{\imath}-\hat{\jmath}),
      m-\eta\hat{\imath}}(u)T_0(u)K^-_0(u)\hat{T}_0(u)\phi^{0}_{m,
      m-\eta\hat{\imath}}(-u)\no\\
  &=&\hspace{-0.32truecm}\sum_{\mu,\nu}\tilde{\phi}^{0}_{m-\eta(\hat{\imath}-\hat{\jmath}),
      m-\eta\hat{\imath}}(u)T_0(u)
      \phi^0_{l-\eta(\hat{\nu}-\hat{\mu}),l-\eta\hat{\nu}}(u)
      \tilde{\phi}^0_{l-\eta(\hat{\nu}-\hat{\mu}),l-\eta\hat{\nu}}(u)\no\\
  &&\quad\times K^-_0(u)\phi^0_{l,l-\eta\hat{\nu}}(-u)
      \bar{\phi}^0_{l,l-\eta\hat{\nu}}(-u)
      \hat{T}_0(u)\phi^{0}_{m,
      m-\eta\hat{\imath}}(-u)\no\\
  &=&\hspace{-0.22truecm}\sum_{\mu,\nu}T(m-\eta\hat{\imath},l-\eta\hat{\nu}|u)^j_{\mu}
       \K(l|u)^{\mu}_{\nu}S(m,l|u)_i^{{\nu}}\no\\
  &\stackrel{{\rm def}}{=}& \T^-(m,l|u)^j_i,
\eea where the face-type K-matrix $\K(l|u)^{\mu}_{\nu}$ is given
by
\bea
 \K(l|u)^{\mu}_{\nu}=\tilde{\phi}^0_{l-\eta(\hat{\nu}-\hat{\mu}),l-\eta\hat{\nu}}(u)
    K^-_0(u)\phi^0_{l,l-\eta\hat{\nu}}(-u).\label{K-1}
\eea Similarly, we have
\bea
 \T^+(m|u)^j_i&=&\prod_{k\neq j}\frac{\s(m_{jk})}{\s\lt(m_{jk}-\eta\rt)}
       \sum_{\mu,\nu}T(l-\eta\hat{\mu},m-\eta\hat{\jmath}|u)_i^{\nu}
       \tilde{\K}(l|u)^{\mu}_{\nu}S(l,m|u)^j_{{\mu}}\no\\
&\stackrel{{\rm
       def}}{=}&\T^+(l,m|u)^j_i
\eea with
\bea
 \tilde{\K}(l|u)^{\mu}_{\nu}=\bar{\phi}^0_{l,l-\eta\hat{\mu}}(-u)
    K^+_0(u)\phi^0_{l-\eta(\hat{\mu}-\hat{\nu}),l-\eta\hat{\mu}}(u).
    \label{K-2}
\eea Thanks to the fact that when $l=\l$ the corresponding
face-type K-matrices $\K(\l|u)$  (\ref{K-1}) and
$\tilde{\K}(\l|u)$  (\ref{K-2}) become diagonal ones (\ref{K-F-3})
and (\ref{K-F-4}), we have \bea
 \T^-(m,\l|u)^j_i\hspace{-0.22truecm}&=&\hspace{-0.22truecm}
       \sum_{\mu}T(m-\eta\hat{\imath},\l-\eta\hat{\mu}|u)^j_{\mu}
       k(\l|u)_{\mu}S(m,\l|u)_i^{{\mu}},\label{New-M-F-1}\\
 \T^+(\l,m|u)^j_i\hspace{-0.22truecm}&=&\hspace{-0.22truecm}
\prod_{k\neq j}\frac{\s(m_{jk})}{\s\lt(m_{jk}-\eta\rt)}
       \sum_{\mu}T(\l-\eta\hat{\mu},m-\eta\hat{\jmath}|u)_i^{\mu}
       \tilde{k}(\l|u)_{\mu}S(\l,m|u)^j_{{\mu}},\label{New-M-F-2}
 \eea where the functions $k(\l|u)_{\mu}$ and $\tilde{k}(\l|u)_{\mu}$ are given by
 (\ref{K-F-3}) and (\ref{K-F-4}) respectively. The relation
(\ref{Crossing-operator}) implies that one can further express
$\T^{\pm}(m|u)^j_i$ in terms of only $T(m,l|u)^j_i$. Here we
present the results for the pseudo-particle creation operators
$\T^{-}(m|u)^2_1$ in (\ref{Bethe-state-2}) and $\T^{+}(m|u)^1_2$
in (\ref{Bethe-state-1}):
\bea
 \T^-(m|u)^2_1\hspace{-0.22truecm}&=&\hspace{-0.22truecm}
      \T^-(m,\l|u)^2_1=\frac{\s(m_{21})}{\s(\l_{21})}\prod_{k=1}^N
      \frac{\s(u+z_k)}{\s(u+z_k+\eta)}\no\\
      &&\,\times\lt\{
      \frac{\s(\l_1+\xi-u)}{\s(\l_1+\xi+u)}
      T(m+\eta\hat{2},\l+\eta\hat{2}|u)^2_1
      T(m,\l|-u-\eta)^2_2\rt.\no\\
 &&\,\quad-\lt.
      \frac{\s(\l_2+\xi-u)}{\s(\l_2+\xi+u)}
      T(m+\eta\hat{2},\l+\eta\hat{1}|u)^2_2
      T(m,\l|-u-\eta)^2_1\rt\},\label{Expression-1}\\
 \T^+(m|u)^1_2\hspace{-0.22truecm}&=&\hspace{-0.22truecm}
      \T^+(\l,m|u)^1_2=\prod_{k=1}^N
      \frac{\s(u+z_k)}{\s(u+z_k+\eta)}\no\\
      &&\,\times\lt\{
      \frac{\s(\l_{12}\hspace{-0.08truecm}-\hspace{-0.08truecm}\eta)
      \s(\l_1\hspace{-0.08truecm}+\hspace{-0.08truecm}\bar{\xi}\hspace{-0.08truecm}+\hspace{-0.08truecm}u
      \hspace{-0.08truecm}+\hspace{-0.08truecm}\eta)}
      {\s(m_{12}\hspace{-0.12truecm}-\hspace{-0.12truecm}\eta)
      \s(\l_1\hspace{-0.12truecm}+\hspace{-0.12truecm}\bar{\xi}\hspace{-0.12truecm}-\hspace{-0.12truecm}u
      \hspace{-0.12truecm}-\hspace{-0.12truecm}\eta)}
      T(\l\hspace{-0.12truecm}+\hspace{-0.12truecm}\eta\hat{2},m\hspace{-0.12truecm}+\hspace{-0.12truecm}\eta\hat{2}|u)^1_2
      T(\l,m|\hspace{-0.12truecm}-\hspace{-0.12truecm}u\hspace{-0.12truecm}-\hspace{-0.12truecm}\eta)^2_2\rt.\no\\
 &&\,\quad\hspace{-0.12truecm}-\hspace{-0.12truecm}\lt.
      \frac{\s(\l_{21}\hspace{-0.12truecm}-\hspace{-0.12truecm}\eta)
      \s(\l_2\hspace{-0.12truecm}+\hspace{-0.12truecm}\bar{\xi}\hspace{-0.12truecm}+\hspace{-0.12truecm}u
      \hspace{-0.12truecm}+\hspace{-0.12truecm}\eta)}
      {\s(m_{21}\hspace{-0.12truecm}+\hspace{-0.12truecm}\eta)
      \s(\l_2\hspace{-0.12truecm}+\hspace{-0.12truecm}\bar{\xi}\hspace{-0.12truecm}-\hspace{-0.12truecm}u
      \hspace{-0.12truecm}-\hspace{-0.12truecm}\eta)}
      T(\l\hspace{-0.12truecm}+\hspace{-0.12truecm}\eta\hat{1},m
      \hspace{-0.12truecm}+\hspace{-0.12truecm}\eta\hat{2}|u)^2_2
      T(\l,m|\hspace{-0.12truecm}-\hspace{-0.12truecm}u\hspace{-0.12truecm}
      -\hspace{-0.12truecm}\eta)^1_2\rt\}.\no\\
 &&\label{Expression-2}
\eea Similar to (\ref{crospendence-1}), we have the
correspondence,
\bea
 &&\T^-(m,l|u)^2_1|i_1,\ldots,i_N\rangle^{m}_{l}\,\longleftrightarrow\,
   \T^-_{F}(m,l|u)^2_1|i_1,\ldots,i_N\rangle,\\
 &&\T^+(m,l|u)^1_2|i_1,\ldots,i_N\rangle^{m}_{l}\,\longleftrightarrow\,
   \T^+_{F}(m,l|u)^1_2|i_1,\ldots,i_N\rangle.
\eea It follows from (\ref{Expression-1}) and (\ref{Expression-2})
that the face-type double-row monodromy matrix elements
$\T^-_F(m,\l|u)^2_1$ and $\T^+_F(\l,m|u)^1_2$ can be expressed in
terms of the face-type one-row monodromy matrix elements
$T_F(m,l|u)^i_j $ (\ref{Monodromy-face-2}) by
\bea
 &&\T^-_F(m,\l|u)^2_1=
 \frac{\s(m_{21})}{\s(\l_{21})}\prod_{k=1}^N
      \frac{\s(u+z_k)}{\s(u+z_k+\eta)}\no\\
 &&\,\quad\times\lt\{
      \frac{\s(\l_1+\xi-u)}{\s(\l_1+\xi+u)}
      T_F(m,\l|u)^2_1
      T_F(m+\eta\hat{2},\l+\eta\hat{2}|-u-\eta)^2_2\rt.\no\\
 &&\,\qquad-\lt.
      \frac{\s(\l_2+\xi-u)}{\s(\l_2+\xi+u)}
      T_F(m+2\eta\hat{2},\l|u)^2_2
      T_F(m+\eta\hat{1},\l+\eta\hat{1}|-u-\eta)^2_1\rt\},\label{Expression-3}\\
 &&\T^+_F(\l,m|u)^1_2=\prod_{k=1}^N
      \frac{\s(u+z_k)}{\s(u+z_k+\eta)}\no\\
 &&\,\quad\times\lt\{
      \frac{\s(\l_{12}\hspace{-0.08truecm}-\hspace{-0.08truecm}\eta)
      \s(\l_1\hspace{-0.08truecm}+\hspace{-0.08truecm}\bar{\xi}
      \hspace{-0.08truecm}+\hspace{-0.08truecm}u\hspace{-0.08truecm}+\hspace{-0.08truecm}\eta)}
      {\s(m_{12}\hspace{-0.08truecm}-\hspace{-0.08truecm}\eta)
      \s(\l_1\hspace{-0.08truecm}+\hspace{-0.08truecm}\bar{\xi}\hspace{-0.08truecm}-
      \hspace{-0.08truecm}u\hspace{-0.08truecm}-\hspace{-0.08truecm}\eta)}
      T_F(\l\hspace{-0.08truecm}+\hspace{-0.08truecm}2\eta\hat{2},m
      \hspace{-0.08truecm}+\hspace{-0.08truecm}2\eta\hat{2}|u)^1_2
      T_F(\l\hspace{-0.08truecm}+\hspace{-0.08truecm}\eta\hat{2},m
      \hspace{-0.08truecm}+\hspace{-0.08truecm}\eta\hat{2}|
      \hspace{-0.08truecm}-\hspace{-0.08truecm}u
      \hspace{-0.08truecm}-\hspace{-0.08truecm}\eta)^2_2\rt.\no\\
 &&\,\qquad-\lt.
      \frac{\s(\l_{21}\hspace{-0.08truecm}-\hspace{-0.08truecm}\eta)
      \s(\l_2\hspace{-0.08truecm}+\hspace{-0.08truecm}\bar{\xi}
      \hspace{-0.08truecm}+\hspace{-0.08truecm}u
      \hspace{-0.08truecm}+\hspace{-0.08truecm}\eta)}
      {\s(m_{21}\hspace{-0.08truecm}+\hspace{-0.08truecm}\eta)
      \s(\l_2\hspace{-0.08truecm}+\hspace{-0.08truecm}\bar{\xi}
      \hspace{-0.08truecm}-\hspace{-0.08truecm}u
      \hspace{-0.08truecm}-\hspace{-0.08truecm}\eta)}
      T_F(\l,m\hspace{-0.08truecm}+\hspace{-0.08truecm}2\eta\hat{2}|u)^2_2
      T_F(\l\hspace{-0.08truecm}+\hspace{-0.08truecm}\eta\hat{2},m
      \hspace{-0.08truecm}+\hspace{-0.08truecm}\eta\hat{2}|
      \hspace{-0.08truecm}-\hspace{-0.08truecm}u\hspace{-0.08truecm}-\hspace{-0.08truecm}
      \eta)^1_2\rt\}.\no\\
 &&\label{Expression-4}
\eea In the derivation of the above equations we have used the
identity $\hat{1}+\hat{2}=0$. Finally, we obtain the face versions
$|v^{(1)}_1,\cdots,v^{(1)}_M\rangle^{(I)}_F$ and
$|v^{(1)}_1,\cdots,v^{(1)}_M\rangle^{(II)}_F$ of the two sets of
Bethe states (\ref{Bethe-state-1}) and (\ref{Bethe-state-2}),
\bea
 &&|v^{(1)}_1,\cdots,v^{(1)}_M\rangle^{(I)}_F=
     \T^+_F(\l,\l\hspace{-0.08truecm}+\hspace{-0.08truecm}2\eta\hat{1}
     |v^{(1)}_1)^1_2\cdots
   \T^+_F(\l,\l\hspace{-0.08truecm}+\hspace{-0.08truecm}2M\eta\hat{1}
   |v^{(1)}_M)^1_2|1,\ldots,1\rangle,
   \label{Bethe-state-3}\\
 &&|v^{(2)}_1,\cdots,v^{(2)}_M\rangle^{(II)}_F =
   \T^-_F(\l\hspace{-0.08truecm}-\hspace{-0.08truecm}2\eta\hat{2},\l
   |v^{(2)}_1)^2_1
   \cdots
   \T^-_F(\l\hspace{-0.08truecm}-\hspace{-0.08truecm}2M\eta\hat{2},\l
   |v^{(2)}_M)^2_1
   |2,\ldots,2\rangle.
   \label{Bethe-state-4}
\eea

In the next section with the help of the Drinfeld twist (or
factorizing F-matrix) in the face picture of the eight-vertex SOS model \cite{Alb00}, we
study expressions of the pseudo-particle creation operators $\T^{\pm}_F$ given by
(\ref{Expression-3}) and (\ref{Expression-4}) in the F-basis. We find that in the magic F-basis
these operators indeed  take  completely symmetric and polarization free forms simultaneously.


\section{ F-basis}
\label{F} \setcounter{equation}{0}

In this section, after briefly reviewing the result \cite{Alb00} about the Drinfeld twist
\cite{Dri83} (factorizing F-matrix) of the eight-vertex SOS model, we
obtain the explicit expression of the double rows
monodromy operator $\T^{\mp}_F(m,\l|u)^2_1$ given by (\ref{Expression-3}) and (\ref{Expression-4}) in
the F-basis provided by the F-matrix.

\subsection{Factorizing Drinfeld twist $F$}
Let $ \mathcal{S}_N$ be the permutation group over indices
$1,\ldots,N$ and $\{s_i|i=1,\ldots,N-1\}$ be the set of
elementary permutations in $\mathcal{S}_N$. For each elementary
permutation $s_i$, we introduce the associated operator
$R^{s_i}_{1\ldots N}$ on the quantum space
\bea
  R^{s_i}_{1\ldots N}(l)\equiv R^{s_i}(l)=R_{i,i+1}
    (z_i-z_{i+1}|l-\eta\sum_{k=1}^{i-1}h^{(k)}),\label{Fundamental-R-operator}
\eea where $l$ is a generic vector in $V$. For any $s,\,s'\in
\mathcal{S}_N$, operator $R^{ss'}_{1\ldots N}$ associated with
$ss'$ satisfies the following composition law
 \cite{Mai00,Alb00,Yan06-1}:
\bea
  R_{1\ldots N}^{ss'}(l)=R^{s'}_{s(1\ldots
  N)}(l)\,R^{s}_{1\ldots N}(l).\label{Rule}
\eea Let $s$ be decomposed in a minimal way in terms of
elementary permutations,
\bea
  s=s_{\b_1}\ldots s_{\b_p}, \label{decomposition}
\eea where $\b_i=1,\ldots, N-1$ and the positive integer $p$ is
the length of $s$. The composition law (\ref{Rule}) enables one
to obtain  operator $R^{s}_{1\ldots N}$ associated with each
$s\in\mathcal{S}_N $. The dynamical quantum Yang-Baxter equation
(\ref{MYBE}), the weight conservation condition (\ref{Conservation})
and the unitary condition (\ref{Unitary}) guarantee the uniqueness of
$R^{s}_{1\ldots N}$. Moreover, one may check that
$R^{s}_{1\ldots N}$ satisfies the following exchange relation
with the face type one-row monodromy matrix
(\ref{Monodromy-face-1}) \bea
  R^{s}_{1\ldots N}(l)T^F_{0,1\ldots N}(l|u)=T^F_{0,s(1\ldots N)}(l|u)
    R^{s}_{1\ldots N}(l-\eta h^{(0)}),\quad\quad \forall s\in
    \mathcal{S}_N.\label{Exchang-Face-1}
\eea

Now, we construct the face-type Drinfeld twist $F_{1\ldots
N}(l)\equiv F_{1\ldots N}(l;z_1,\ldots,z_N)$ \footnote{In this
paper, we adopt the convention: $F_{s(1\ldots N)}(l)\equiv
F_{s(1\ldots N)}(l;z_{s(1)},\ldots,z_{s(N)})$.} on the $N$-fold
tensor product space $V^{\otimes N}$, which  satisfies the
following three properties \cite{Alb00,Yan06-1}:
\bea
 &&{\rm I.\,\,\,\,lower-triangularity;}\\
 &&{\rm II.\,\,\, non-degeneracy;}\\
 &&{\rm III.\,factorizing \, property}:\,\,
 R^{s}_{1\ldots N}(l)\hspace{-0.08truecm}=\hspace{-0.08truecm}
    F^{-1}_{s(1\ldots N)}(l)F_{1\ldots N}(l), \,\,
 \forall s\in  \mathcal{S}_N.\label{Factorizing}
\eea Substituting (\ref{Factorizing}) into the exchange relation
(\ref{Exchang-Face-1}), we have
\bea
 F^{-1}_{s(1\ldots N)}(l)F_{1\ldots N}(l)T^F_{0,1\ldots N}(l|u)=
   T^F_{0,s(1\ldots N)}(l|u)F^{-1}_{s(1\ldots N)}(l-\eta h^{(0)})
   F_{1\ldots N}(l-\eta h^{(0)}).
\eea Equivalently,
\bea
 F_{1\ldots N}(l)T^F_{0,1\ldots N}(l|u)F^{-1}_{1\ldots N}(l-\eta h^{(0)})
   =F_{s(1\ldots N)}(l)T^F_{0,s(1\ldots N)}(l|u)
   F^{-1}_{s(1\ldots N)}(l-\eta h^{(0)}).\label{Invariant}
\eea Let us introduce the twisted monodromy matrix
$\tilde{T}^F_{0,1\ldots N}(l|u)$ by \bea
 \tilde{T}^F_{0,1\ldots N}(l|u)&=&
  F_{1\ldots N}(l)T^F_{0,1\ldots N}(l|u)F^{-1}_{1\ldots N}(l-\eta
  h^{(0)})\no\\
  &=&\lt(\begin{array}{ll}\tilde{T}_F(l|u)^1_1&\tilde{T}_F(l|u)^1_2
  \\\tilde{T}_F(l|u)^2_1&
   \tilde{T}_F(l|u)^2_2\end{array}\rt).\label{Twisted-Mon-F}
\eea Then (\ref{Invariant}) implies that the twisted monodromy
matrix is symmetric under $\mathcal{S}_N$, namely, \bea
 \tilde{T}^F_{0,1\ldots N}(l|u)=\tilde{T}^F_{0,s(1\ldots
 N)}(l|u), \quad \forall s\in \mathcal{S}_N.
\eea

Define the F-matrix:
\bea
  F_{1\ldots N}(l)=\sum_{s\in
     \mathcal{S}_N}\sum^2_{\{\a_j\}=1}\hspace{-0.22truecm}{}^*\,\,\,\,
     \prod_{i=1}^NP^{s(i)}_{\a_{s(i)}}
     \,R^{s}_{1\ldots N}(l),\label{F-matrix}
\eea where $P^i_{\a}$ is the embedding of the project operator
$P_{\a}$ in the $i^{{\rm th}}$ space with matric elements
$(P_{\a})_{kl}=\d_{kl}\d_{k\a}$. The sum $\sum^*$ in
(\ref{F-matrix}) is over all non-decreasing sequences of the
labels $\a_{s(i)}$:
\bea
  && \a_{s(i+1)}\geq \a_{s(i)}\quad {\rm if}\quad s(i+1)>s(i),\\
  && \a_{s(i+1)}> \a_{s(i)}\quad {\rm if}\quad
  s(i+1)<s(i).\label{Condition}
\eea From (\ref{Condition}), $F_{1\ldots N}(l)$ obviously is a
lower-triangular matrix. Moreover, the F-matrix is non-degenerate
because  all its diagonal elements are non-zero. It was shown in \cite{Alb00} that
the F-matrix also satisfies the factorizing property (\ref{Factorizing}).

\subsection{Completely symmetric  representations}
In the F-basis provided by the F-matrix (\ref{F-matrix}),  the
twisted operators $\tilde{T}_F(l|u)^j_i$ defined by
(\ref{Twisted-Mon-F}) become polarization free \cite{Alb00}. Here we present the
results relevant for our purpose
\bea
 &&\tilde{T}_F(l|u)^2_2=\frac{\s(l_{21}-\eta)}{\s\lt(l_{21}-\eta+
     \eta\langle H,\e_1\rangle\rt)}\otimes_{i}
     \lt(\begin{array}{ll}\frac{\s(u-z_i)}{\s(u-z_i+\eta)}&\\
     &1\end{array}\rt)_{(i)},\\[6pt]
 &&\tilde{T}_F(l|u)^2_1=\sum_{i=1}^N\frac{\s(\eta)
     \s(u\hspace{-0.08truecm}-\hspace{-0.08truecm}z_i
     \hspace{-0.08truecm}+\hspace{-0.08truecm}l_{12})}
     {\s(u\hspace{-0.08truecm}-\hspace{-0.08truecm}z_i
     \hspace{-0.08truecm}+\hspace{-0.08truecm}\eta)\s(l_{12})} E_{12}^i\otimes_{j\neq i}
     \lt(\begin{array}{ll}\frac{\s(u-z_j)\s(z_i-z_j+\eta)}{\s(u-z_j+\eta)\s(z_i-z_j)}&\\
     &1\end{array}\rt)_{(j)},\\[6pt]
 &&\tilde{T}_F(l|u)^1_2=\frac{\s(l_{21}\hspace{-0.08truecm}-\hspace{-0.08truecm}\eta)}
     {\s(l_{21}\hspace{-0.08truecm}+\hspace{-0.08truecm}
     \eta\langle H,\e_1\hspace{-0.08truecm}-\hspace{-0.08truecm}\e_2\rangle)}
     \hspace{-0.08truecm}
     \sum_{i=1}^N\frac{\s(\eta)\s(u\hspace{-0.08truecm}-\hspace{-0.08truecm}z_i
     \hspace{-0.08truecm}+\hspace{-0.08truecm}l_{21}\hspace{-0.08truecm}
     +\hspace{-0.08truecm}\eta\hspace{-0.08truecm}
     +\hspace{-0.08truecm}\eta\langle H,\e_1
     \hspace{-0.08truecm}-\hspace{-0.08truecm}\e_2\rangle)}
     {\s(u\hspace{-0.08truecm}-\hspace{-0.08truecm}z_i\hspace{-0.08truecm}+\hspace{-0.08truecm}\eta)
     \s(l_{21}\hspace{-0.08truecm}+\hspace{-0.08truecm}\eta
     \hspace{-0.08truecm}+\hspace{-0.08truecm}\eta\langle
     H,\e_1\hspace{-0.08truecm}-\hspace{-0.08truecm}\e_2\rangle)}\no\\
 &&\quad\quad\quad\quad\quad\quad
     \times E_{21}^i\otimes_{j\neq i} \lt( \begin{array}{ll}
     \frac{\s(u-z_j)}{\s(u-z_j+\eta)}&\\
     &\frac{\s(z_j-z_i+\eta)}{\s(z_j-z_i)}\end{array}
     \rt)_{(j)},
\eea where $H=\sum_{k=1}^N h^{(k)}$. Applying  the above operators
to the arbitrary  state $|i_1,\ldots,i_N\rangle$ given by
(\ref{Vector-V}), we have
\bea
 &&\tilde{T}_F(m,l|u)^2_2=\frac{\s(l_{21}-\eta)}{\s\lt(l_{2}-m_1-\eta\rt)}
     \otimes_{i}
     \lt(\begin{array}{ll}\frac{\s(u-z_i)}{\s(u-z_i+\eta)}&\\
     &1\end{array}\rt)_{(i)},\\[6pt]
 &&\tilde{T}_F(m,l|u)^2_1=\sum_{i=1}^N
     \frac{\s(\eta)
     \s(u-z_i+l_{12})}{\s(u-z_i+\eta)\s(l_{12})}\no\\
 &&\quad\quad\quad\quad\quad\quad
     \times    E_{12}^i \otimes_{j\neq i}
     \lt(\begin{array}{ll}\frac{\s(u-z_j)\s(z_i-z_j+\eta)}{\s(u-z_j+\eta)\s(z_i-z_j)}&\\
     &1\end{array}\rt)_{(j)},\\[6pt]
 &&\tilde{T}_F(m,l|u)^1_2=\frac{\s(l_{21}-\eta)}
     {\s(m_{21}-2\eta)}
     \sum_{i=1}^N\frac{\s(\eta)\s(u-z_i+m_{21}-\eta)}
     {\s(u-z_i+\eta)\s(m_{21}-\eta)}\no\\
 &&\quad\quad\quad\quad\quad\quad
     \times E_{21}^i\otimes_{j\neq i} \lt( \begin{array}{ll}
     \frac{\s(u-z_j)}{\s(u-z_j+\eta)}&\\
     &\frac{\s(z_j-z_i+\eta)}{\s(z_j-z_i)}\end{array}
     \rt)_{(j)}.
\eea With the help of the Riemann identity (\ref{identity}), we find that the two pseudo-particle creation
operators (\ref{Expression-3}) and (\ref{Expression-4}) in the
F-basis simultaneously have the following completely symmetric
polarization free forms:
\bea
 &&\tilde{\T}^-_F(m,\l|u)^2_1=\frac{\s(m_{12})}{\s(m_1-\l_2)}
   \prod_{k=1}^N\frac{\s(u+z_k)}{\s(u+z_k+\eta)}\no\\
  &&\quad\quad\times \sum_{i=1}^N\frac{\s(\l_1+\xi-z_i)\s(\l_2+\xi+z_i)\s(2u) \s(\eta)}
   {\s(\l_1+\xi+u)\s(\l_2+\xi+u)\s(u-z_i+\eta)\s(u+z_i)}\no\\[6pt]
  &&\quad\quad\quad\quad\quad\quad \times
   E_{12}^i\otimes_{j\neq i}\lt(\begin{array}{ll}
   \frac{\s(u-z_j)\s(u+z_j+\eta)\s(z_i-z_j+\eta)}
   {\s(u-z_j+\eta)\s(u+z_j)\s(z_i-z_j)}&\\
   &1\end{array}\rt)_{(j)},\label{Creation-operator-1}\\[6pt]
 &&\tilde{\T}^+_F(\l,m|u)^1_2=\frac{\s(m_{21}+\eta)}{\s(m_2-\l_1)}
   \prod_{k=1}^N\frac{\s(u+z_k)}{\s(u+z_k+\eta)}\no\\
  &&\quad\quad\times \sum_{i=1}^N
   \hspace{-0.08truecm}
   \frac{\s(\l_2\hspace{-0.08truecm}+\hspace{-0.08truecm}\bar{\xi}
   \hspace{-0.08truecm}-\hspace{-0.08truecm}z_i)
   \s(\l_1\hspace{-0.08truecm}+\hspace{-0.08truecm}\bar{\xi}
   \hspace{-0.08truecm}+\hspace{-0.08truecm}z_i)
   \s(2u\hspace{-0.08truecm}+\hspace{-0.08truecm}2\eta) \s(\eta)}
   {\s(\l_1\hspace{-0.08truecm}+\hspace{-0.08truecm}\bar{\xi}
   \hspace{-0.08truecm}-\hspace{-0.08truecm}u
   \hspace{-0.08truecm}-\hspace{-0.08truecm}\eta)
   \s(\l_2\hspace{-0.08truecm}+\hspace{-0.08truecm}\bar{\xi}
   \hspace{-0.08truecm}-\hspace{-0.08truecm}u
   \hspace{-0.08truecm}-\hspace{-0.08truecm}\eta)
   \s(u\hspace{-0.08truecm}+\hspace{-0.08truecm}z_i)
   \s(u\hspace{-0.08truecm}-\hspace{-0.08truecm}z_i
   \hspace{-0.08truecm}+\hspace{-0.08truecm}\eta)}\no\\[6pt]
  &&\quad\quad\quad\quad\quad\quad \times
   E_{21}^i\otimes_{j\neq i}\lt(\begin{array}{ll}
   \frac{\s(u-z_j)\s(u+z_j+\eta)}
   {\s(u-z_j+\eta)\s(u+z_j)}&\\
   &\frac{\s(z_j-z_i+\eta)}{\s(z_j-z_i)}\end{array}\rt)_{(j)}.\label{Creation-operator-2}
\eea The very polarization free form (\ref{Creation-operator-1}) of
$\tilde{\T}^-_F(m,\l|u)^2_1$ enabled the authors in \cite{Yan11} to succeed in
obtaining  a single determinant representation of the domain wall partition function
of the eight-vertex model with a non-diagonal reflection end.


\section{Bethe states in F-basis}
\label{BF} \setcounter{equation}{0}
Now let us evaluate  the two sets of Bethe states
(\ref{Bethe-state-3}) and (\ref{Bethe-state-4}) in the F-basis (or
the twisted Bethe states)
\bea
 \overline{|v^{(1)}_1,\cdots,v^{(1)}_M\rangle}^{(I)}_F&=&F_{1\ldots N}(\l)
          \,|v^{(1)}_1,\cdots,v^{(1)}_M\rangle^{(I)}_F,\label{Twisted-Bethe-state-1}\\
 \overline{|v^{(2)}_1,\cdots,v^{(2)}_M\rangle}^{(II)}_F&=&F_{1\ldots N}(\l)
          \,|v^{(2)}_1,\cdots,v^{(2)}_M\rangle^{(II)}_F.\label{Twisted-Bethe-state-2}
\eea Since $|1,\ldots,1\rangle$ and $|2,\ldots,2\rangle$ are
invariant under the action of the F-matrix $F_{1\ldots N}(l)$
(\ref{F-matrix}), namely, \bea
 F_{1\ldots N}(l)\,|i,\ldots,i\rangle=|i,\ldots,i\rangle,\quad
 i=1,2,\no
\eea we have \bea
 &&\overline{|v^{(1)}_1,\cdots,v^{(1)}_M\rangle}^{(I)}_F=
     \tilde{\T}^+_F(\l,\l\hspace{-0.08truecm}+\hspace{-0.08truecm}2\eta\hat{1}
     |v^{(1)}_1)^1_2\cdots
     \tilde{\T}^+_F(\l,\l\hspace{-0.08truecm}+\hspace{-0.08truecm}2M\eta\hat{1}
     |v^{(1)}_M)^1_2|1,\ldots,1\rangle,
   \label{Twisted-Bethe-state-3}\\
 &&\overline{|v^{(2)}_1,\cdots,v^{(2)}_M\rangle}^{(II)}_F =
   \tilde{\T}^-_F(\l\hspace{-0.08truecm}-\hspace{-0.08truecm}2\eta\hat{2},\l
   |v^{(2)}_1)^2_1
   \cdots
   \tilde{\T}^-_F(\l\hspace{-0.08truecm}-\hspace{-0.08truecm}2M\eta\hat{2},\l
   |v^{(2)}_M)^2_1
   |2,\ldots,2\rangle.
   \label{Twisted-Bethe-state-4}
\eea Thanks to the polarization free representations
(\ref{Creation-operator-1}) and (\ref{Creation-operator-2}) of the
pseudo-particle creation operators,  we can obtain completely
symmetric expressions of the two sets of Bethe sates in the
F-basis:
\bea
 \overline{|v^{(1)}_1,\cdots,v^{(1)}_M\rangle}^{(I)}_F&=&\prod_{k=1}^M
     \lt\{ \frac{\s(\l_{12}-\eta+2k\eta)}{\s(\l_{12}+k\eta)}
     \prod_{n=1}^{N}\frac{\s(v^{(1)}_k-z_n)}{\s(v^{(1)}_k-z_n+\eta)}\rt\}\no\\
 &&\quad\quad \times \sum_{i_1<i_2\ldots<i_M}B^{(I)}_M\lt(\{v^{(1)}_{\a}\}|\{z_{i_n}\}\rt)
     E^{i_1}_{21}\ldots
     E^{i_M}_{21}\,|1,\ldots,1\rangle,\label{BA-state-1}\\
 \overline{|v^{(2)}_1,\cdots,v^{(2)}_M\rangle}^{(II)}_F&=&\prod_{k=1}^M
     \lt\{ \frac{\s(\l_{12}+2k\eta)}{\s(\l_{12}+k\eta)}
     \prod_{n=1}^{N}\frac{\s(v^{(2)}_k+z_n)}{\s(v^{(2)}_k+z_n+\eta)}\rt\}\no\\
 &&\quad\quad \times \sum_{i_1<i_2\ldots<i_M}B^{(II)}_M\lt(\{v^{(2)}_{\a}\}|\{z_{i_n}\}\rt)
     E^{i_1}_{12}\ldots
     E^{i_M}_{12}\,|2,\ldots,2\rangle.\label{BA-state-2}
\eea Here the functions $B^{(I)}_M\lt(\{v_{\a}\}|\{z_{i}\}\rt)$
and $B^{(II)}_M\lt(\{v_{\a}\}|\{z_{i}\}\rt)$ are given by
\bea
 &&B^{(I)}_M\lt(\{v_{\a}\}|\{z_{i}\}\rt)=\prod_{n=1}^M\prod_{i=1}^M
   \frac{\s(v_n-z_{i}+\eta)\s(v_n+z_{i})}{\s(v_n-z_{i})\s(v_n+z_{i}+\eta)}\no\\
 &&\,\quad\times \sum_{s\in\mathcal{S}_M}\prod_{n=1}^M\hspace{-0.08truecm}
   \lt\{\frac{\s(\l_2\hspace{-0.08truecm}+\hspace{-0.08truecm}\bar{\xi}
   \hspace{-0.08truecm}-\hspace{-0.08truecm}z_{s(n)})
   \s(\l_1\hspace{-0.08truecm}+\hspace{-0.08truecm}\bar{\xi}
   \hspace{-0.08truecm}+\hspace{-0.08truecm}z_{s(n)})
   \s(2v_n\hspace{-0.08truecm}+\hspace{-0.08truecm}2\eta)\s(\eta)}
   {\s(\l_2\hspace{-0.08truecm}+\hspace{-0.08truecm}\bar{\xi}
   \hspace{-0.08truecm}-\hspace{-0.08truecm}v_n
   \hspace{-0.08truecm}-\hspace{-0.08truecm}\eta)
   \s(\l_1\hspace{-0.08truecm}+\hspace{-0.08truecm}\bar{\xi}
   \hspace{-0.08truecm}-\hspace{-0.08truecm}v_n
   \hspace{-0.08truecm}-\hspace{-0.08truecm}\eta)
   \s(v_n\hspace{-0.08truecm}+\hspace{-0.08truecm}z_{s(n)})
   \s(v_n\hspace{-0.08truecm}-\hspace{-0.08truecm}z_{s(n)}
   \hspace{-0.08truecm}+\hspace{-0.08truecm}\eta)}
   \rt.\no\\
 &&\,\quad\quad \times \prod_{k>n}^M\lt.
   \frac{\s(v_k\hspace{-0.08truecm}-\hspace{-0.08truecm}z_{s(n)})
    \s(v_k\hspace{-0.08truecm}+\hspace{-0.08truecm}z_{s(n)}
    \hspace{-0.08truecm}+\hspace{-0.08truecm}\eta)
    \s(z_{s(k)}\hspace{-0.08truecm}-\hspace{-0.08truecm}z_{s(n)}
    \hspace{-0.08truecm}+\hspace{-0.08truecm}\eta)}
   {\s(v_k\hspace{-0.08truecm}-\hspace{-0.08truecm}z_{s(n)}
   \hspace{-0.08truecm}+\hspace{-0.08truecm}\eta)
   \s(v_k\hspace{-0.08truecm}+\hspace{-0.08truecm}z_{s(n)})
   \s(z_{s(k)}\hspace{-0.08truecm}-\hspace{-0.08truecm}z_{s(n)})}
   \rt\},\label{Function-B-1}\\[6pt]
 &&B^{(II)}_M\lt(\{v_{\a}\}|\{z_{i}\}\rt)=\hspace{-0.18truecm}
   \sum_{s\in\mathcal{S}_M}\prod_{n=1}^M\hspace{-0.08truecm}
   \lt\{\frac{\s(\l_1\hspace{-0.08truecm}+\hspace{-0.08truecm}\xi
   \hspace{-0.08truecm}-\hspace{-0.08truecm}z_{s(n)})
   \s(\l_2\hspace{-0.08truecm}+\hspace{-0.08truecm}\xi
   \hspace{-0.08truecm}+\hspace{-0.08truecm}z_{s(n)})
   \s(2v_n)\s(\eta)}
   {\s(\l_1\hspace{-0.08truecm}+\hspace{-0.08truecm}\xi
   \hspace{-0.08truecm}+\hspace{-0.08truecm}v_n)
   \s(\l_2\hspace{-0.08truecm}+\hspace{-0.08truecm}\xi
   \hspace{-0.08truecm}+\hspace{-0.08truecm}v_n)
   \s(v_n\hspace{-0.08truecm}-\hspace{-0.08truecm}z_{s(n)}
   \hspace{-0.08truecm}+\hspace{-0.08truecm}\eta)
   \s(v_n\hspace{-0.08truecm}+\hspace{-0.08truecm}z_{s(n)})}
   \rt.\no\\
   &&\,\quad\quad \times \prod_{k>n}^M\lt.
   \frac{\s(v_n-z_{s(k)}) \s(v_n+z_{s(k)}+\eta) \s(z_{s(n)}-z_{s(k)}+\eta)}
   {\s(v_n-z_{s(k)}+\eta) \s(v_n+z_{s(k)}) \s(z_{s(n)}-z_{s(k)})}
   \rt\}.\label{Function-B-2}
\eea From the expressions (\ref{Function-B-1}) and (\ref{Function-B-2}), it is
easy to check that these two functions $B^{(i)}_M\lt(\{v_{\a}\}|\{z_{i}\}\rt)$ are
symmetric functions of $\{v_a\}$ and $\{z_i\}$ separatively. Following the method of \cite{Yan10-1,Yan11},
we can further express the functions in terms of some single determinant respectively as follows:
\bea
B^{(I)}_M\lt(\{v_{\a}\}|\{z_{i}\}\rt)=
   \frac{\prod_{\a=1}^M\prod_{i=1}^M\s(v_{\a}+z_i)\s(v_{\a}-z_i+\eta)
   \,{\rm det}{\cal N}^{(I)}(\{v_{\a}\};\{z_i\})}
  {\prod_{\a>\b}\s(v_{\a}\hspace{-0.1truecm}-\hspace{-0.1truecm}
  v_{\b})\s(v_{\a}\hspace{-0.1truecm}+\hspace{-0.1truecm}v_{\b}
  \hspace{-0.1truecm}+\hspace{-0.1truecm}\eta)\prod_{k<l}
  \s(z_l\hspace{-0.1truecm}-\hspace{-0.1truecm}z_k)\s(z_l\hspace{-0.1truecm}+\hspace{-0.1truecm}z_k)},
  \label{partition-1}\\
B^{(II)}_M\lt(\{v_{\a}\}|\{z_{i}\}\rt)=
   \frac{\prod_{\a=1}^M\prod_{i=1}^M\s(v_{\a}-z_i)\s(v_{\a}+z_i+\eta)
   \,{\rm det}{\cal N}^{(II)}(\{v_{\a}\};\{z_i\})}
  {\prod_{\a>\b}\s(v_{\a}\hspace{-0.1truecm}-\hspace{-0.1truecm}
  v_{\b})\s(v_{\a}\hspace{-0.1truecm}+\hspace{-0.1truecm}v_{\b}
  \hspace{-0.1truecm}+\hspace{-0.1truecm}\eta)\prod_{k<l}
  \s(z_k\hspace{-0.1truecm}-\hspace{-0.1truecm}z_l)\s(z_k\hspace{-0.1truecm}+\hspace{-0.1truecm}z_l)},
  \label{partition-2}
\eea where the $M\times M$ matrices  ${\cal N}^{(i)}(\{v_{\a}\};\{z_i\})$ are given by
\bea
{\cal N}^{(I)}(\{v_{\a}\};\{z_i\})_{\a,j}&=&
  \frac{\s(\eta)\s(\l_2+\bar{\xi}-z_j)}
  {\s(v_{\a}-z_j)\s(v_{\a}+z_j+\eta)
  \s(\l_2+\bar{\xi}-v_{\a}-\eta)}\no\\[8pt]
  &&\times \frac{\s(\l_1+\bar{\xi}+z_j)\s(2v_{\a}+2\eta)}
  {\s(\l_1+\bar{\xi}-v_{\a}-\eta)
  \s(v_{\a}-z_j+\eta)
  \s(v_{\a}+z_j)},\label{Matrix-1}\\
{\cal N}^{(II)}(\{v_{\a}\};\{z_i\})_{\a,j}&=&
  \frac{\s(\eta)\s(\l_1+\xi-z_j)}
  {\s(v_{\a}-z_j)\s(v_{\a}+z_j+\eta)
  \s(\l_1+\xi+v_{\a})}\no\\[8pt]
  &&\times \frac{\s(\l_2+\xi+z_j)\s(2v_{\a})}
  {\s(\l_2+\xi+v_{\a})
  \s(v_{\a}-z_j+\eta)
  \s(v_{\a}+z_j)}.\label{Matrix-2}
\eea

We remark that if the parameters $\{v^{(1)}_k\}$ (or
$\{v^{(2)}_k\}$) do not satisfy the associated Bethe ansatz
equations (\ref{BA-D-1})(or (\ref{BA-D-2})), the corresponding
twisted states (\ref{Twisted-Bethe-state-3}) and
(\ref{Twisted-Bethe-state-4}) become off-shell Bethe states. These
off-shell Bethe states can still be expressed in the same forms as
those of (\ref{BA-state-1})-(\ref{Matrix-2}) (but the
corresponding parameters are not necessarily the roots of the
Bethe ansatz equations).


\section{ Conclusions}
\label{C} \setcounter{equation}{0}

We have studied the explicit expressions of the Bethe states of
the open XYZ chain with non-diagonal boundary terms (where the
non-diagonal K-matrices $K^{\pm}(u)$ are given by
(\ref{K-matrix-2-1}) and (\ref{K-matrix-6})) by using the Drinfeld twist or factorizing
F-matrix for the eight-vertex SOS model. It is found that in
the F-basis the pseudo-particle creation operators, which generate
the two sets of the eigenstates of the model, simultaneously
take the completely symmetric and polarization free forms
(\ref{Creation-operator-1}) and (\ref{Creation-operator-2}). This
allows us to obtain the explicit and completely symmetric explicit
expressions (\ref{BA-state-1}) and (\ref{BA-state-2}) of the two
sets of (off-shell) Bethe states. Moreover, the coefficients
$B^{(i)}_M\lt(\{v_{\a}\}|\{z_{i}\}\rt)$ in these  expressions
can be expressed in terms of a single determinant (\ref{partition-1})-(\ref{Matrix-2})
respectively. Such a single determinant representation makes it feasible to derive the
determinant representations of  scalar
products of the Bethe states for the open XYZ chain with
non-diagonal boundary terms specified by the non-diagonal
K-matrices $K^{\pm}(u)$ (\ref{K-matrix-2-1}) and
(\ref{K-matrix-6}).

\section*{Acknowledgements}
The financial supports from  the National Natural Science
Foundation of China (Grant Nos. 11075126 and 11031005) and  Australian Research
Council are gratefully acknowledged.


\end{document}